\documentclass[times, twocolumn]{aastex631mod}

\usepackage{hyperref}
\hypersetup{
  colorlinks = true,
  citecolor = blue,
  linkcolor = blue,
  urlcolor = blue}

\usepackage{graphicx}
\usepackage{subfigure}
\usepackage{amssymb}
\usepackage{epstopdf}
\usepackage{natbib}
\usepackage{color}
\usepackage{subfigure}
\usepackage{ulem}
\usepackage{multirow}
\def\gtrsim{\lower 2pt \hbox{$\, \buildrel {\scriptstyle >}\over
{\scriptstyle \sim}\,$}}
\def\lesssim{\lower 2pt \hbox{$\, \buildrel {\scriptstyle <}\over
{\scriptstyle \sim}\,$}}

\def\xmm{{XMM-Newton}}
\def\chandra{{Chandra}}

\def\phcm{\hbox{photon cm$^{-2}$ s$^{-1}$ }}

\def\cmcu{\hbox{cm$^{-3}$}}
\def\kmps{\hbox{km $\rm{s^{-1}}$}}

\def\fexvii{Fe~{\sc xvii}}

\def\hi{H~{\sc i}}

\def\neix{Ne~{\sc ix}}
\def\nex{Ne~{\sc x}}

\def\oviii{O~{\sc viii}}
\def\ovii{O~{\sc vii}}

\def\mgxi{Mg~{\sc xi}}

\def\six{Si~{\sc x}}

\def\heii{He~{\sc ii}}

\usepackage{soul} 

\shorttitle{CX Emission in the Disk of M51}
\shortauthors{Zhang et al.}

\begin{document}

\title{X-ray Spectroscopic Evidence of Charge Exchange Emission in the Disk of M51}


\author[0000-0002-8329-6606]{Shuinai Zhang}
\affiliation{Purple Mountain Observatory, Chinese Academy of Sciences, Nanjing 210033, China}
\affiliation{Key Laboratory of Dark Matter and Space Astronomy, Chinese Academy of Sciences, Nanjing, China}

\author[0000-0002-9279-4041]{ Q. Daniel Wang}
\affiliation{Department of Astronomy, University of Massachusetts, Amherst, Massachusetts, USA}

\author[0000-0002-5456-0447]{Wei Sun}
\affiliation{Purple Mountain Observatory, Chinese Academy of Sciences, Nanjing 210033, China}

\author[0000-0001-5938-5070]{Min Long}
\affiliation{Department of Computer Science, Boise State University, Boise, Idaho, USA}

\author[0000-0002-2541-2373]{Jia Sun}
\affiliation{Purple Mountain Observatory, Chinese Academy of Sciences, Nanjing 210033, China}

\author[0000-0001-7500-0660]{Li Ji}
\affiliation{Purple Mountain Observatory, Chinese Academy of Sciences, Nanjing 210033, China}
\affiliation{Key Laboratory of Dark Matter and Space Astronomy, Chinese Academy of Sciences, Nanjing, China}

\correspondingauthor{Shuinai Zhang}
\email{snzhang@pmo.ac.cn}



\begin{abstract}
In the disks of spiral galaxies, diffuse soft X-ray emission is known to be strongly correlated with star-forming regions.
However, this emission is not simply from a thermal-equilibrium plasma and its origin remains greatly unclear.  
In this work, we present an X-ray spectroscopic analysis of the emission from the northern hot spot; a region with enhanced star-formation off the nucleus of M51.
Based on the high spectral resolution data from \xmm/RGS observations, we unambiguously detect a high $G$ ratio ($3.2^{+6.9}_{-1.5}$) of the \ovii\ He$\alpha$ triplet. 
This high $G$ ratio is also spatially confirmed by oxygen emission-line maps from the same data.
A physical model consisting of a thermal plasma and its charge exchange (CX) with neutral cool gas gives a good explanation for the $G$ ratio and the entire RGS spectra. This model also gives a satisfactory characterization of the complementary \chandra~ACIS-S data, which enables a direct imaging of the diffuse emission, tracing the hot plasma across the galaxy. 
The hot plasma has a similar characteristic temperature of $\sim0.34$ keV and an approximately solar metallicity. 
The CX contributes $\sim$50\% to the diffuse emission in the 0.4--1.8~keV~band, suggesting an effective hot/cool gas interface area about five times the geometric area of the M51 disk. 
Therefore, the CX appears to play a major role in the soft X-ray production and may be used as a powerful tool to probe the interface astrophysics, important for  studying galactic ecosystems.  
\end{abstract}

\keywords{X-ray observatories; Spiral galaxies; High resolution spectroscopy; Stellar feedback; Diffuse radiation}

\section{Introduction}

Diffuse hot plasma at temperature $\gtrsim10^6\,\rm K$ in galactic disks is believed to be chiefly due to stellar feedback in forms of  fast winds and supernova explosions of massive stars \citep[e.g.,][]{McKee77, Li13}. Such plasma, if sufficiently energetic, can reshape the interstellar medium (ISM) and transport the feedback energy and chemically enriched matter into galactic halos. However, how the plasma may be characterized is still under debate, largely because its interplay with the cool ISM remains uncertain. 

Currently, the diagnostics of the diffuse hot plasma are basically based on X-ray spectral observations with CCD energy resolution ($\sim$100 eV), which is not refined enough to measure many individual emission lines in the soft X-ray band. As a result, spectral analysis of diffusion X-ray emission is typically based on ad hoc models with varied complexities, assuming an optically thin thermal plasma under the collisional ionization equilibrium (CIE). One often needs two-temperature CIE components peaking at $\lesssim$0.2 keV and $\gtrsim$0.5 keV to fit the plasma contribution to a spectrum \citep[e.g.,][]{Owen09, Mineo12, Li13}. When its counting statistics are good, additional components are often needed \citep[e.g., in the face-on galaxy M101;][]{Kuntz10}. Alternatively, the spectra may be fitted with a continuous temperature distribution of the emission measure \citep[e.g.,][]{Kuntz10}. Indeed,  the X-ray spectra of the 30 Doradus nebula in the Large Magellanic Cloud \citep{Cheng21} and the galactic disk of M83 \citep{Wang21} can be well characterized by a log-normal temperature distribution.

Multiple scenarios have been proposed to explain such multitemperature structures of hot plasma in the ISM. The two-temperature structures, for example, may be caused by a mixture of  hot bubbles in a galactic disk with different temperatures \citep{Doane04}. In other studies, the hotter thermal component may be emission from dwarf stars and supernovae in the interarm regions, while the cooler one may represent clumpy star-forming regions in its disk and also accumulated gas in a shallow halo; the two components are not necessarily in pressure equilibrium \citep[e.g.,][]{Warwick07}. But the reality is likely more complicated, since one would also expect the hot outflows from a disk into a galactic halo \citep{Hodges-Kluck18} and their adiabatic cooling \citep{Breitschwerdt99}. 

To better understand the origin of the diffuse X-ray emission, as well as the physical and chemical properties of the hot plasma in galaxies, high-resolution spectroscopic data are needed. The Reflection Grating Spectrometer (RGS) on board \xmm\ has much higher spectral resolution ($\sim$2 eV), $\sim$50 times better than the CCD energy resolution, and can be used to better constrain hot plasma properties or even reveal different radiative mechanisms other than the CIE thermal emission. For example, RGS observations have been used to show that the charge exchange (CX) at interfaces between hot plasma and neutral gas may contribute a significant portion to the diffuse X-ray emission in the starburst galaxy M82 \citep{Zhang14} and that this situation is likely prevalent in nearby galaxies \citep{Wang12}. One indicator of the existence of CX is the high $G$ ratio of \ovii~He$\alpha$ triplet lines  [$=(f + i)/r\gtrsim1.4$, where the resonance ({\it r}) line is at 21.602 \AA\ (or 574 eV), the intercombination ({\it i}) line is at 21.804 \AA\ (or 569 eV), and the forbidden ({\it f}) line is at 22.098 \AA\ (or 561 eV)]. In these studies, the RGS spectra tend to be dominated by strong emissions from galactic central regions, which complicates the interpretation of the observed high $G$ ratios, which could, in principle, arise from past active galactic nuclei \citep[AGN;][]{Zhang19} or jet interactions with circumnuclear gas \citep{Yang20}, for example.

Here we present an investigation of diffuse X-ray emission from a well-isolated region in the galactic disk of M51  (also known as M51a, NGC 5194, or the Whirlpool galaxy), based on \xmm/RGS observations and \chandra/ACIS observations. M51 is a nearly face-on galaxy at the distance of 8.58~Mpc \citep{McQuinn16} and has a mean surface rate of star formation  $\sim 0.015~M_\odot$~yr$^{-1}$~kpc$^{-2}$ \citep{Calzetti05}. The galaxy is normally classified to be ``quiescently star-forming.'' Nevertheless, there is a particular striking recent star-forming region at the northern east part of the M51 disk, as evidenced by the presence of multiple massive young stellar clusters \citep{Kaleida10}, as well as enhanced far UV and H$\alpha$ emission \citep{Thilker00, Owen09}. We call this region as the northern hot spot (NHS). The intense star formation there might be triggered by the collapse of molecular clouds due to the tidal compression of the companion galaxy NGC 5195 \citep{Dobbs10}. The NHS has an angular size of  $\sim 2'$ ($\sim5$ kpc) and is $\sim 2'$ away from the galactic nucleus or the X-ray-bright companion galaxy NGC~5195. This relative compactness and isolation of the NHS makes it possible to study its X-ray emission effectively with the slitless spectroscopic capability of the RGS. We use archival \xmm/RGS observations covering the NHS, complemented by high spatial resolution X-ray observations from \chandra/ACIS, to probe the origin and properties of the X-ray emission observed over much of the galactic disk, as well as the NHS.

The Paper is organized as follows. We describe the \xmm~and \chandra~X-ray observations and data reduction, as well the use of a complementary \hi~image, in Section \ref{sec:data}. The X-ray data analysis and results are presented in Section \ref{sec:analysis}. We discuss the implications of our results in Section \ref{sec:discuss}, and summarize the work and our conclusions in Section \ref{sec:conclusion}.

\section{Observations and Data Reduction}
\label{sec:data}

\subsection{\xmm/RGS Data}

Our study relies chiefly on RGS observations. Table~\ref{tab:log} lists five \xmm/RGS observations that cover the NHS of the M51 disk. Our reduction process follows the standard procedure provided by the science analysis system (SAS; version19). This includes the removal of time intervals with strong background flares, and to produce the RGS spectra and the auxiliary files, using the pipeline command ``{\tt rgsproc}.'' The total effective exposure is 354 ks. However, our spectral analysis of the NHS uses only the first four observations.  Observation ID (ObsID) 0852030101 was used only in the emission-line mapping of \oviii~and \ovii~{\it f}, because the observation had a bad column right at the key \ovii~{\it r} line.

\begin{deluxetable}{cccc} 
\tablecaption{RGS observation log of M51.}
\tablecolumns{5}
\tablewidth{0pt}
\tablehead{
\colhead{ObsID}  & \colhead{Date}                  &  \colhead{$t_{\rm eff}$}       & \colhead{P.A.}\\
\colhead{}            &  \colhead{(yy-mm-dd)} &  \colhead{(ks)}                     & \colhead{($^{\circ}$)}
}
\startdata
0824450901  & 2018-05-13    &  77   &  326  \\  
0830191401  & 2018-05-25    &  83   &  326   \\  
0830191501  & 2018-06-13   &  62   &   308  \\ 
0830191601  & 2018-06-15   &  62   &  306   \\  
0852030101  & 2019-07-11   &  70   &   287 \\
\enddata
\label{tab:log}
\tablecomments{ $t_{\rm eff}$ denotes the effective exposures after filtering out time intervals with strong background flares, while P.A. stands for the position angles of the RGS observations.}
\end{deluxetable}

\begin{figure*} 
\subfigure[]{
\begin{minipage}[t]{2.3in}
\centering 
         \includegraphics[angle=0,width=\textwidth]{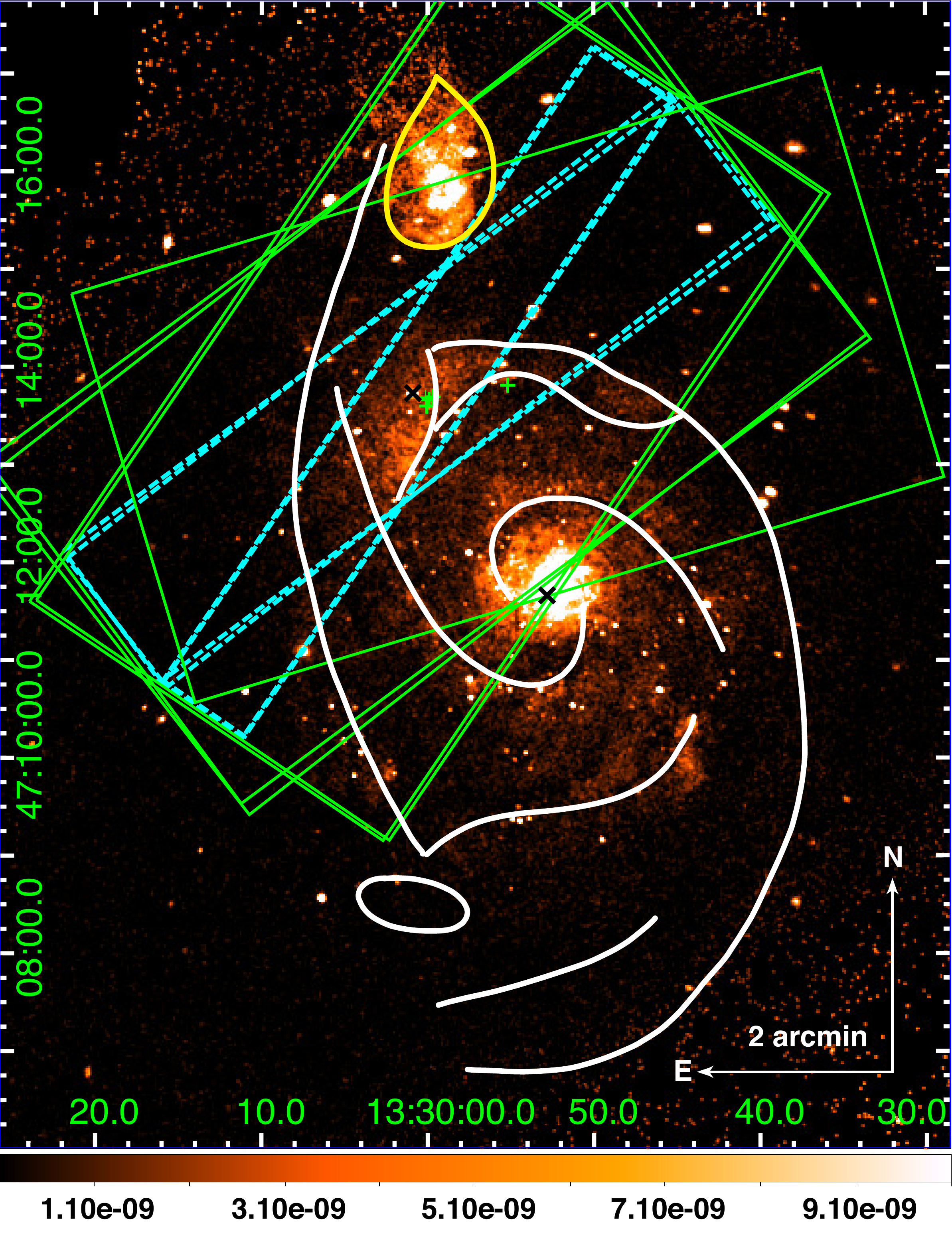}  
\end{minipage}%
}%
\subfigure[]{
\begin{minipage}[t]{2.3in}
\centering
        \includegraphics[angle=0,width=\textwidth]{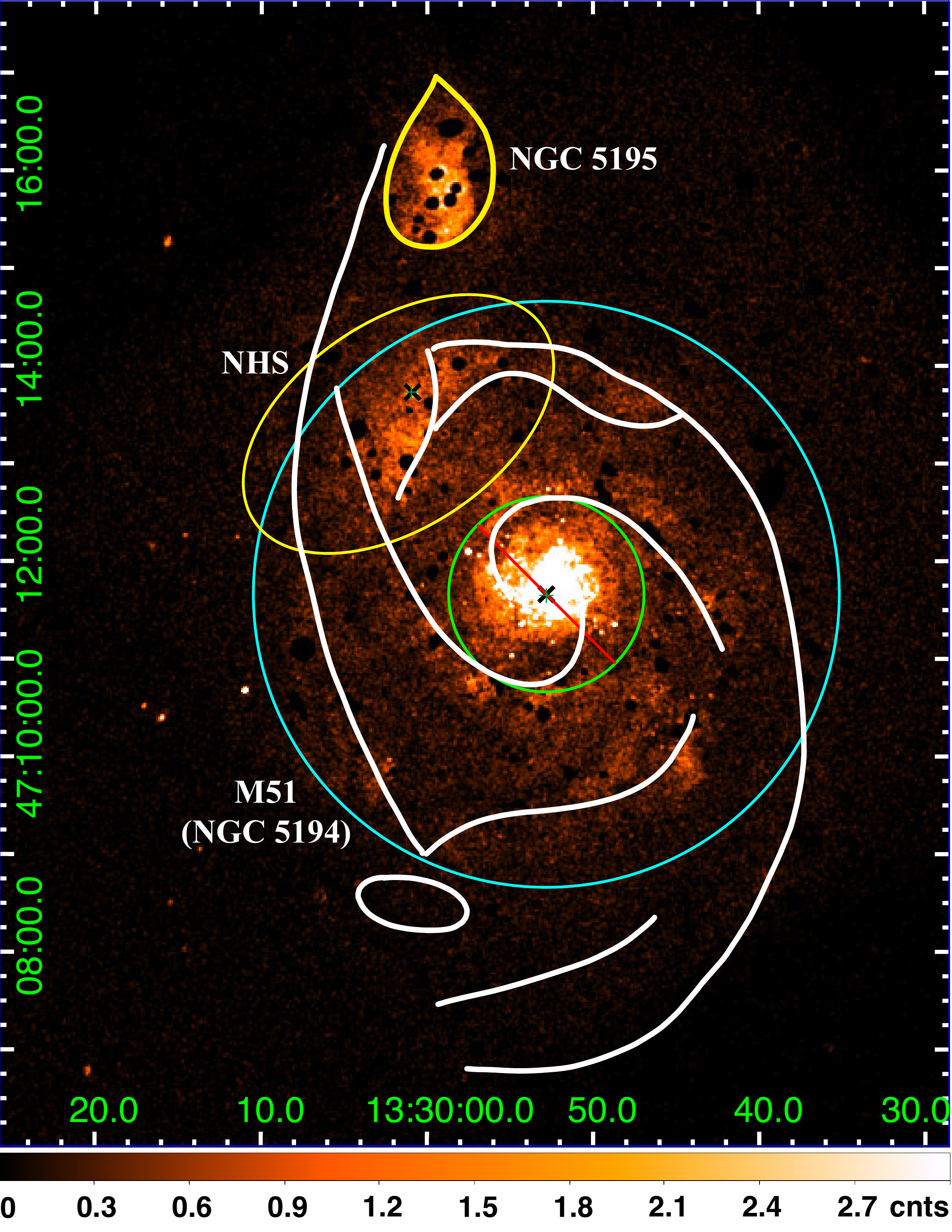}
\end{minipage}%
}%
\subfigure[]{
\begin{minipage}[t]{2.3in}
\centering 
       \includegraphics[angle=0,width=\textwidth]{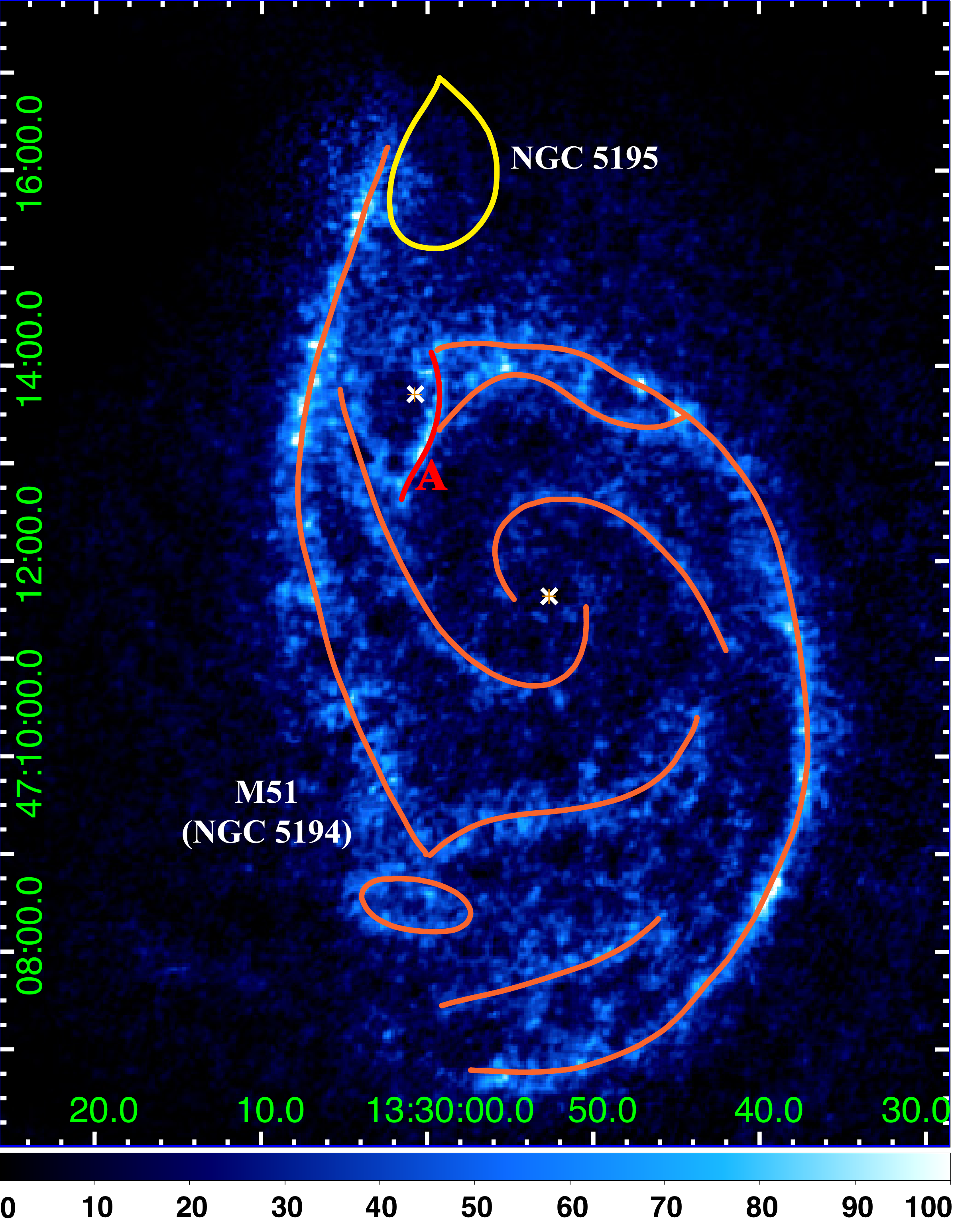}
\end{minipage}%
}%
 \caption{
(a) \chandra/ACIS-S intensity image of M51 in the 0.5--1.2 keV band. The green rectangles ($8'\times4\farcm4$) represent the dispersion directions, as well as the widths of the  regions covered by the RGS observations, while their on-axis positions are marked as the green pluses (``+"). The actual spectral extraction regions for the NHS are outlined by the four cyan dashed rectangles with the widths of $1'$ or $1\farcm6$ in the cross-dispersion direction.
(b) The same \chandra~count intensity image of M51, but with detected discrete sources removed. The yellow elliptical region is defined as the NHS region, while the cyan annular region with the elliptical region excluded is defined as the disk region without NHS. 
The image is overlaid with white curves sketching out prominent neutral atomic gas features (mainly spiral arms) as seen in panel (c), where the field is seen in the \hi\ 21 cm line emission in units of $\mathrm{Jy\,beam^{-1}\cdot m\,s^{-1}}$. While the sketch is in pale violet red in panel (c), a vertical arm segment ``A'' is highlighted with the red color. The cross symbols  (``$\times$") in all the three panels mark the galactic center and the reference point for the spectral extraction.
}
\label{fig:regions} 
\end{figure*}

We use the spectral extraction regions as illustrated in Figure~\ref{fig:regions}a, to minimize the contamination from both the nucleus of M51 and the companion galaxy NGC 5195 and set the reference point for ``{\tt rgsproc}''  to R.A.:$13^\mathrm{h}30^\mathrm{m}00\fs890$, decl:$+47^{\circ}13'44\farcs0$. The background spectrum is derived from blank-sky spectral templates, according to the background level indicator as the count rate of the off-axis region on CCD 9. However, the off-axis region contains two AGNs of M51 and NGC 5195, and its count rate is even larger than that of the on-axis NHS region. Therefore, the model background could be overpredicted, and empirically we scale down its flux by 2/3 to get the continuum spectrum between 28 and 29 \AA\ larger than zero. This scaling of the background has little effect on our measurements of emission lines. 

It should be noted that the CCD 7 (10.6--13.8 \AA) in RGS1 and CCD 4 (20.0--24.1 \AA) in RGS2 were nonoperational during the observations. Furthermore, the CCD 6 (13.8--17.1 \AA) in RGS1 contains bad columns and the chip gap right around the prominent \fexvii\ lines, and therefore is also excluded. We combine the RGS1 and the RGS2 spectra separately using the ``{\tt rgscombine}" script. The two combined spectra are grouped with a bin size of 0.05 \AA, which is close to the RGS spectral resolution of 0.07 \AA. A few bins that encounter severe bad columns are ignored, such as the bin at 21.80 \AA.

The RGS data are also used to reconstruct the oxygen emission-line maps that cover the northern part of M51. Each RGS observation contains 1D spatial information in the cross-dispersion direction with a total width of $\sim5'$, where the FWHM of the line spread function is about $22''$ around 20 \AA. In its dispersion direction, the line broadening profile is determined mainly by the spatial extension of the X-ray line emission, following the relation $\Delta\lambda = 0.138\ \Delta\theta $. Therefore, if the Doppler effect can be neglected, as is the case here, the spatial information in the dispersion direction can be deduced from the observed line profile, reaching a spatial resolution of about $30''$. The feasibility of producing RGS line intensity images has been demonstrated previously \citep{Heyden03, Bauer07}. In the present work, we need to combine RGS observations with different dispersion directions to maximize the signal-to-noise ratio (S/N) to map the distributions of the \oviii~Ly$\alpha$ and \ovii~{\it f} or {\it r} lines. These are the strongest lines in the RGS spectra and are relatively isolated from other lines. \ref{sec:procedure} details the procedure for constructing monochromatic intensity maps of the emission lines.

\subsection{\chandra/ACIS-S data}

We use \chandra/ACIS-S data to decompose the diffuse and discrete source contributions and to determine the overall spectral shape and hence the properties of the plasma. We use all the 10 \chandra\ observations taken between 2000 June and 2012 October (ObsID: 354, 1622, 3932, 13812, 13813, 13814, 13815, 13816, 15496, and 15553). The total effective exposure time of these observations is 827~ks, after removing strong background flare periods. We process the data, using the standard \chandra\ interactive analysis of observations (CIAO; version:~4.8). The processing starts from the Level-1 event files through ``{\tt chandra\_repro}.'' After reprojecting the individual event files to the same tangent point, we merge all of them and create the count flux map in the 0.5--1.2 keV band, as shown in Figure~\ref{fig:regions}a \& b. For most of the observations, the separation of the aim points is no more than 1$'$, which makes the point-spread-function of each point source similar. Thus we detect pointlike sources in the 0.3--7.0 keV band, based on the merged data with the CIAO script {\tt wavdetect}, and remove the source ellipses with major and minor axes twice of the 90\% energy encircled values.

We extract the X-ray spectra from the NHS region (the yellow elliptical region in Figure~\ref{fig:regions}b) and the disk region without NHS (the cyan annular region with the yellow NHS region excluded), as well as the corresponding response files, from each observation, using the CIAO script {\tt specextract}. We use {\tt dmextract} and the {\tt blank-sky} datasets to estimate the background spectral contributions. We finally merge the spectral files of individual observations together, using {\tt combine\_spectra}.

\subsection{\hi~Image}
We adopt an \hi~map of M51 from the THINGS project  (Figure~\ref{fig:regions}c; \citealt{Walter08}) to trace neutral gas and its potential interplay with the hot plasma. The integrated \hi~emission also helps us to sketch out the positions of the grand spiral arms and other major cool gas features, which are used as references for multiwavelength comparisons. In particular, a roughly vertical segment of  prominent \hi\ emission, marked as ``A'' in  Figure~\ref{fig:regions}c, appears on the western side of the NHS region. For ease of reference, Figure~\ref{fig:regions}c also includes a yellow curve that encloses an apparent X-ray emission enhancement associated with NGC 5195.

\section{Spectral Analysis and Results}
\label{sec:analysis}

We use \texttt{pyXspec} within {\it Xspec} v12.10\footnote{https://heasarc.gsfc.nasa.gov/docs/xanadu/xspec/} to perform the spectral analysis. The fitting procedure is mainly Markov Chain Monte Carlo based, together with the {\it Cash} statistic. The quoted errors are at the 90\% confidence level for one free parameter case.

\subsection{RGS spectra analysis and results}

We start with the fitting to the RGS spectra of the NHS with a fiducial model, consisting of a power law and a single-temperature APEC plasma, assuming optically thin and CIE. In addition to a Galactic foreground absorption with the column density fixed to the observed \hi\ value of $1.8\times10^{20} \,\rm{cm^{-2}}$ (Table 2; \citealt{Kalberla05}), we include a fitting absorption for the power-law component, representing the contribution from discrete sources in the field, which tend to be embedded in the dense ISM. We assume no X-ray absorption intrinsic to M51 for the plasma component, characterizing the diffuse soft X-ray emission, which arises primarily from an \hi\ cavity with insignificant column density (Figure~\ref{fig:regions}c). The line broadening in the plasma emission is accounted for by the inclusion of the {\it Xspec} convolution model `{\tt rgsxsrc}', together with the 0.5--1.2 keV intensity image of the diffuse X-ray emission of the NHS. We generate the image from the \chandra\ data by filling the holes left from the removal of detected sources via the interpolation from their surrounding source-free areas.

\begin{deluxetable*}{c|ccc|cc}[htb!]
\tablecaption{Model parameters of the spectra}
\tablecolumns{6}
\tablehead{
\colhead{}	  &  \multicolumn{3}{c}{RGS}  & \multicolumn{2}{c}{ACIS-S}   \\
\colhead{}	  &  \multicolumn{3}{c}{NHS} & \colhead{NHS}    & \colhead{Other Arms}
}
\startdata
Local $N_{\rm H}$     &    \multicolumn{5}{c}{$1.8\times10^{20}\,\rm cm^{-2}$ (fixed)}    \\ \hline
 $N_{\rm H\_pl}$ ($\times10^{22}\,\rm cm^{-2}$)        &     $<0.10$     &  $<0.08$   &       $<0.11$        &    $<0.14$  &  $<0.05$     \\ 
$\eta_\mathrm{pl}$\tablenotemark{a}	 [$\times 10^{-4}$] &  $2.1^{+0.4}_{-0.2}$ & $2.3^{+1.7}_{-0.8}$  &     $1.9^{+0.4}_{-0.2}$    & $0.23^{+0.09}_{-0.03}$ &  $0.58^{+0.12}_{-0.01}$ \\
$\Gamma_\mathrm{pl}$ 	        &  $1.9^{+0.4}_{-0.3}$ & $1.7\pm0.2$  &     $1.8^{+0.4}_{-0.3}$    & $1.8^{+0.3}_{-0.1}$      & $3.3^{+0.4}_{-0.1}$  \\ \hline
$\eta_\mathrm{apec}$\tablenotemark{a} [$\times 10^{-5}$]  &    $3.0^{+3.4}_{-1.7}$   & $4.8\pm0.7$ &     ---         & $5.9^{+1.8}_{-0.9}$  &   $10.2^{+1.1}_{-2.2}$   \\
$kT_{APEC}$ (keV)					& $0.25\pm0.02$ 	  & $0.39^{+0.05}_{-0.07}$ 	&  ---    &   $0.34^{+0.02}_{-0.01}$    &    $0.34\pm0.01$  \\  \hline
$\eta_\mathrm{vlntd}$\tablenotemark{a}  [$\times 10^{-5}$] 	& ---      	& ---      	& $10.4^{+5.7}_{-2.2}$	& ---      	& ---      \\ 
$kT_{vlntd}$ (keV)					& ---  & --- 	&  $0.39^{+0.05}_{-0.07}$        &   ---    &    ---  \\
$\sigma_x$    &    ---    &   ---  &   $<0.4$  &  ---  &  ---    \\  \hline
$\eta_\mathrm{acx}$\tablenotemark{a}  [$\times 10^{-5}$] 	& ---                  & $2.6\pm0.5$    &     $1.6^{+1.5}_{-0.9}$	& $4.7\pm1.1$  &    $11.0\pm1.9$   \\
$kT_{ACX}$ (keV)                                         &    ---    &  0.39 (tied) &   $0.25\pm0.10$  &  0.34 (tied)  &  0.34 (tied)   \\  \hline
$\textrm{N} $ 					& $2.8^{+1.0}_{-1.6}$            & $1.0\pm0.7$   &        $1.2^{+0.2}_{-0.6}$  	&   $1.0^{+0.7}_{-0.5}$          &    $0.4^{+0.1}_{-0.2}$   \\
$\textrm{O} $ 					& $1.6^{+2.1}_{-0.8}$            & $0.6^{+0.1}_{-0.3}$ &     $0.7^{+0.1}_{-0.3}$       &  $0.5\pm0.1$                   &    $0.5\pm0.1$  \\
$\textrm{Ne}$ 					& $3.0^{+4.0}_{-1.9}$            & $0.7\pm0.3$   &      $0.8\pm0.2$     &  $1.1\pm0.2$               &   $1.0\pm0.1$  \\
$\textrm{Mg}$ 					& 1 (fixed)                   & 1 (fixed)             & 1 (fixed)              & $1.3^{+0.5}_{-0.2}$    &   $1.1^{+0.2}_{-0.1}$   \\
$\textrm{Fe}$ 					& $1.9^{+2.9}_{-1.1}$   & $0.3\pm0.1$  &   $0.3\pm0.1$     & $0.7\pm0.1$               &  $0.7\pm0.1$   \\
Redshift 				           &  \multicolumn{5}{c}{0.0015 (fixed on the NED value)}     \\
 \hline
$C$-stat./$d.o.f.$	&  725/577 	& 709/576        &   709/574     & 105/78   &  138/80    \\
 \hline
 Thermal flux [$\times 10^{-5}$]\tablenotemark{b}   &   8.6     &   7.3   &   8.1    &     8.0   &    13.4   \\
 CX flux [$\times 10^{-5}$]\tablenotemark{b}           &   --     &   3.5   &   2.8    &    9.6   &   15.1   \\
\enddata
\tablecomments{The columns are: (1) the name of free parameters of the models; (2) parameter values in the fiducial fitting for the NHS with a single {\it APEC} model; (3)  the {\it APEC+ACX} model fitting for the NHS;  (4) the {\it vlntd+ACX} model fitting for the NHS; (5) parameter values in \chandra~CCD spectrum of the NHS; (6) parameter values in \chandra~CCD spectrum of other parts of the galactic disk. Some elements, i.e. N, O, Ne, Mg, and Fe, are allowed to vary, while other metal elements have solar abundances.}
\tablenotetext{a}{Normalization parameters of the {\it power-law} and {\it APEC} models are {\it Xspec} defaults. For the power-law model, the `norm' has the physical meaning of $\rm photons\,keV^{-1}cm^{-2}s^{-1}$ at 1 keV. For the {\it APEC} model and the {\it vlntd} model, the `norm' has a physical meaning of  $\frac{10^{-14}}{4\pi D^2} \int n_{\rm e}n_{\rm H} dV$, where $D$ is the distance to the source (in units of cm), $n_{\rm e}$ and $n_{\rm H}$ are the electron and H densities (cm$^{-3}$), and $V$ is the volume. The `norm' of the {\it ACX} model is similar: $\frac{10^{-10}}{4\pi  D^2} \int n_\mathrm{d} n_\mathrm{r} \mathrm{d} V_\mathrm{i}$, where $n_{\rm d}$ and $n_{\rm r}$ are the number densities of the donors and receivers in the CX process, and $V_\mathrm{i}$ is the volume of an interface layer where CX occurs.}
\tablenotetext{b}{The unabsorbed model flux in the 7--30 \AA\ band and in units of \phcm.}
\label{tab:par}
\end{deluxetable*}

\begin{figure*}
 \centering
\subfigure[]{	\includegraphics[angle=0,width=4.0in]{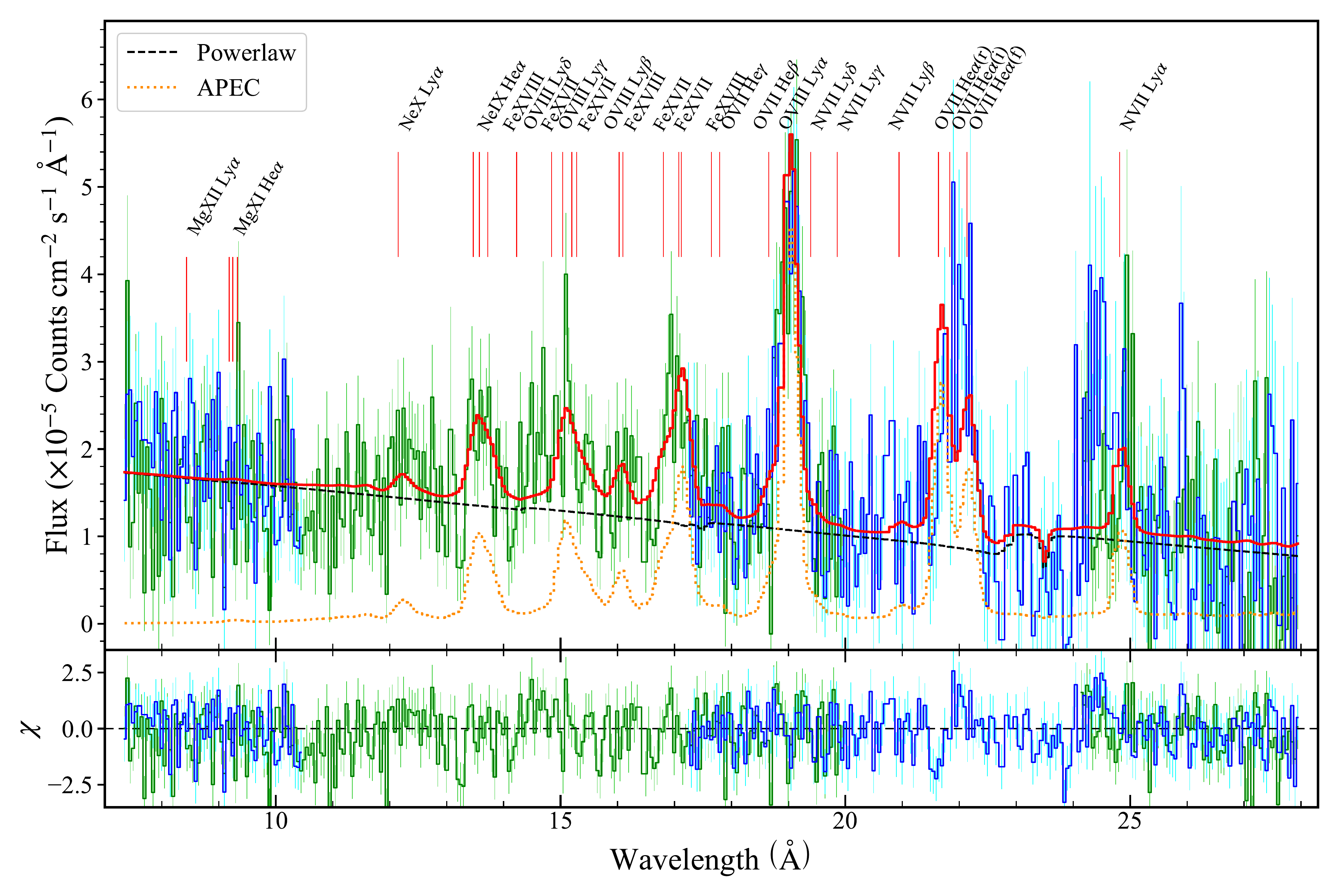} 
	\includegraphics[angle=0,width=2.63in]{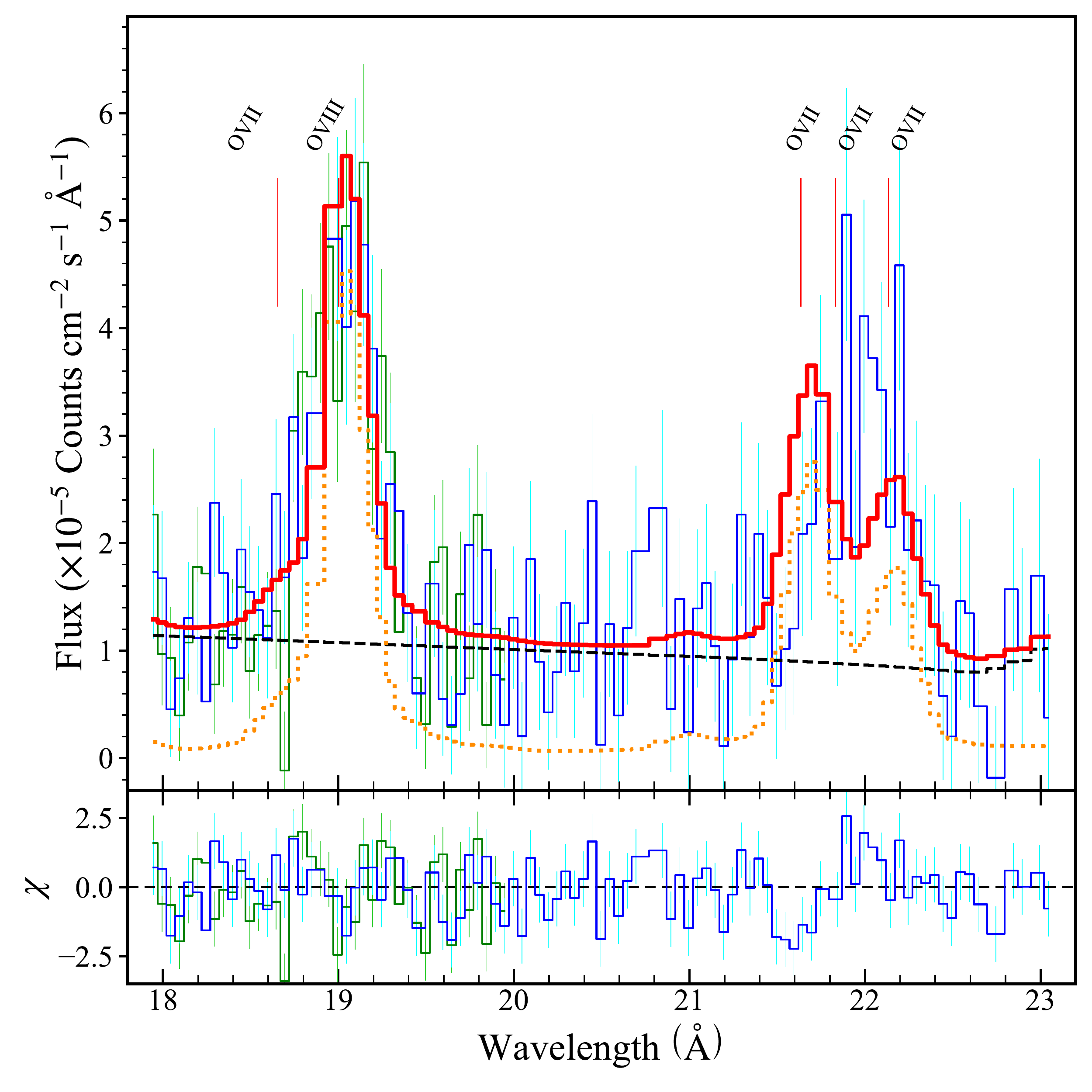}}
\subfigure[]{	\includegraphics[angle=0,width=4.0in]{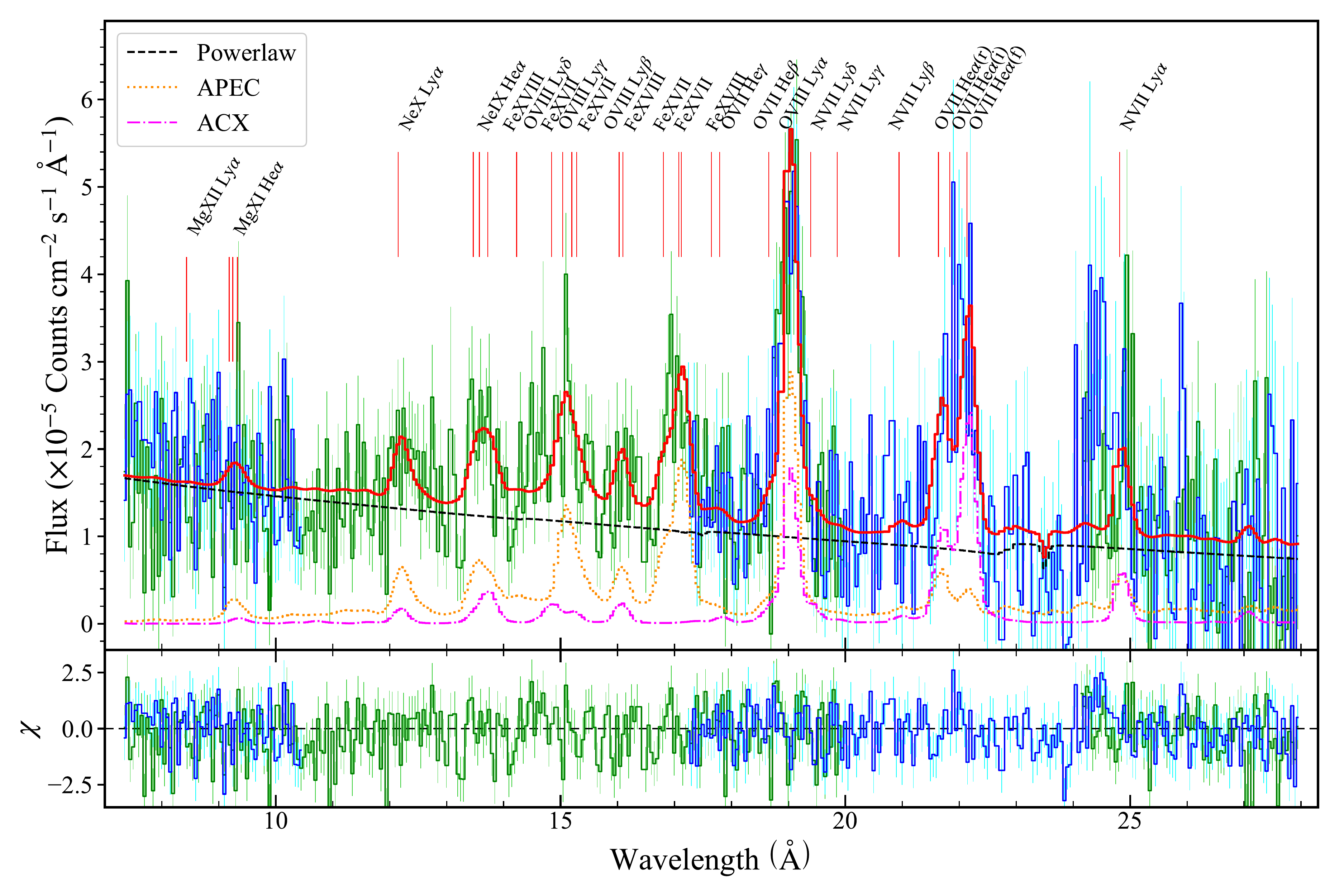} 
	\includegraphics[angle=0,width=2.63in]{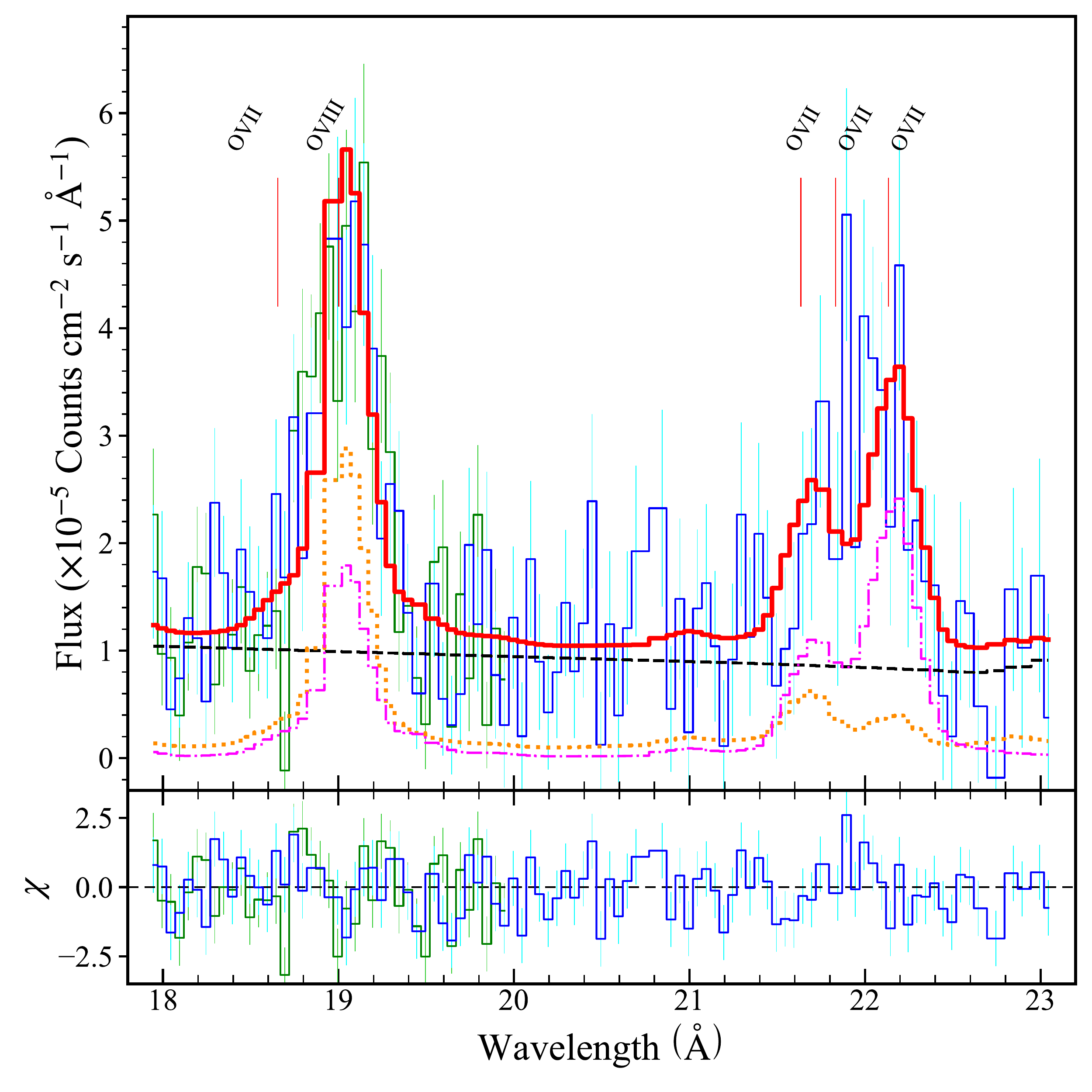}}
\subfigure[]{	\includegraphics[angle=0,width=4.0in]{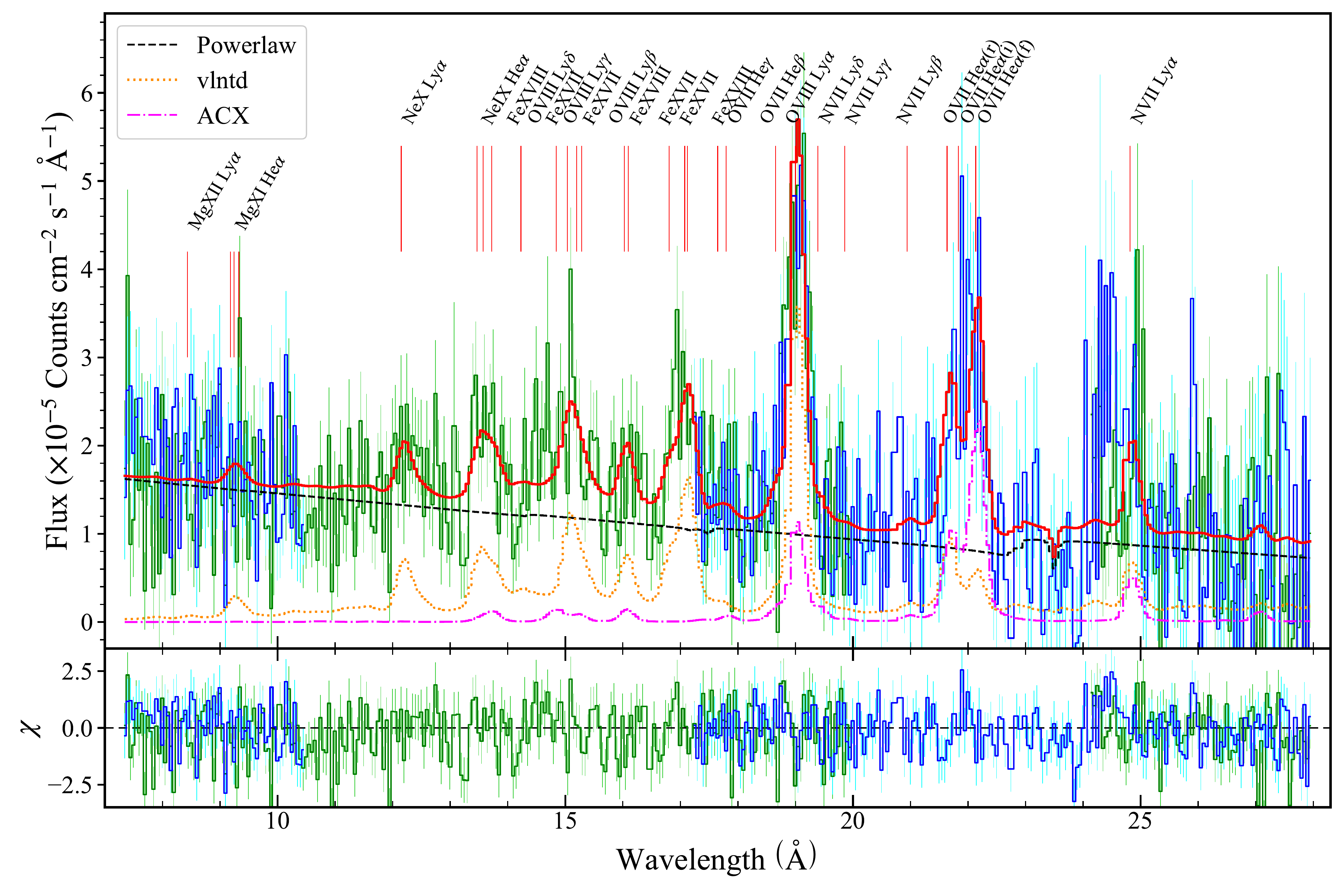} 
	\includegraphics[angle=0,width=2.63in]{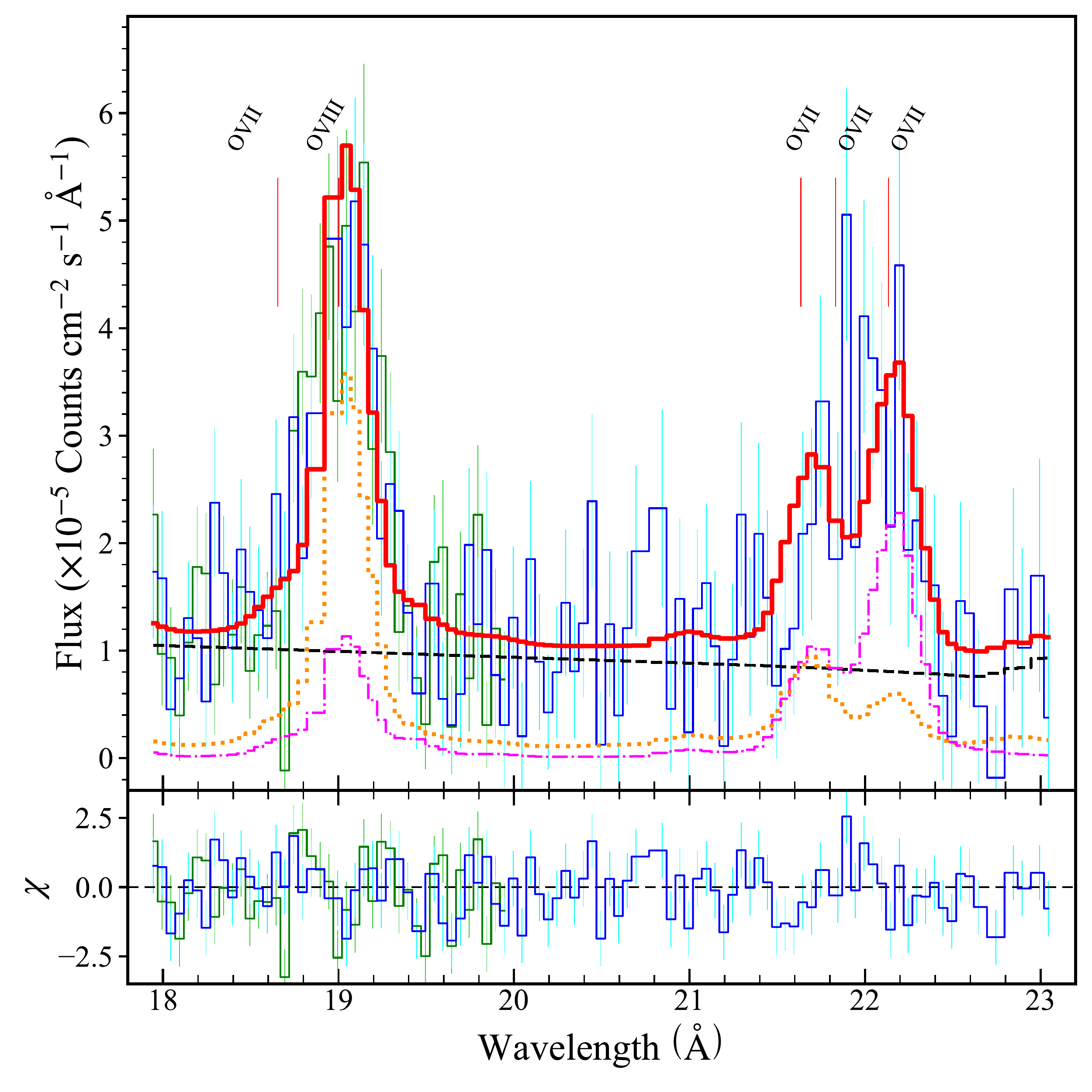}}
\caption{
RGS1 and RGS2 spectra of the NHS (colored coded in blue and green, respectively) in comparison with fitted models: (a) single-temperature {\it APEC} thermal plasma (red line) for the diffuse X-ray emission; (b)  {\it APEC+ACX};
(c) {\it vltnd+ACX} (Table 2). In these panels, while the red histogram represents the total model spectra, the black dashed,  orange dotted, and magenta dotted-dashed curves represent the point source, thermal plasma,  and  CX contributions, respectively. }
\label{fig:apeccx} 
\end{figure*}

Figure~\ref{fig:apeccx}a shows the best-fit fiducial model to the RGS spectra, while the fitted parameters are listed in Table~\ref{tab:par}. Although the model represents well such moderately ionized lines as \oviii, \fexvii, and \neix~lines, the \nex~Ly$\alpha$ lines are not fitted properly even allowing for an abnormally large neon abundance. Also underfitting are the lines from highly ionized magnesium, indicating the need for a higher temperature plasma. More indicative is a clear mismatch between the data and model at the \ovii~He$\alpha$ triplet, where the \ovii~{\it f} line is higher than the model prediction while the {\it r} line is lower. This mismatch strongly suggests that the line emission in the spectra cannot simply arise from a CIE plasma as assumed. In fact, a two-temperature CIE model does not improve the fit to the \ovii~triplet at all.

We calculate the $G$ ratio of the \ovii~He$\alpha$ triplet to diagnose the nature of the diffuse X-ray emission. To do so, we use the `{\tt mdefine}' command in {\it Xspec} to define a \ovii~He$\alpha$ triplet model consisting of three Gaussians, representing the three lines. Their  known energies are fixed to their rest-frame values, while their Gaussian widths to an insignificantly small value (0.001~keV). We further fix the normalization ratio of the weak $i$ line to the $f$ line to 1/4.44, a rather good approximation for an optically thin thermal CIE plasma. As a result, this model has only two fitting parameters: the normalization of the triplet and the $G$ ratio. With the continuum component set to the best-fit fiducial model, the model fits  the RGS1 spectrum well in the 20.8--22.9 \AA\ range. The fitted $G$ ratio ($3.2_{-1.5}^{+6.9}$)  is significantly higher than the value ($\lesssim 1.4$) expected for a CIE plasma. We conclude that this is the strong X-ray spectroscopic evidence for a significant CX contribution to the diffuse soft X-ray emission, while alternative scenarios are discussed in \ref{sec:explanations} and are not favored. 

Accordingly, we systematically include the CX contribution in the modeling of the RGS spectra. Specifically, we account for the CX contribution, using the second version {\it ACX}\footnote{http://www.atomdb.org/CX/} model \citep{Smith12}. This version includes velocity-dependent reaction cross sections  \citep{Mullen2017}, while the previous version used a simple empirical formula for the CX reaction rates. We assume that the encounter velocity between hot ions and cool atoms as 280 \kmps\ (close to the sound speed of a 0.3 keV plasma). The temperature and metal abundances of the hot plasma in the  {\it ACX} and {\it APEC} model components are linked. Therefore, compared to the fiducial model, the new {\it APEC+ACX} model adds only one more fitting parameter,  the normalization of the {\it ACX} model. The quality of the fit to the RGS spectra is improved considerably, reducing the {\it Cash} statistic from 725 to 709 (Table 2). Figure~\ref{fig:apeccx}b shows that the \ovii~He$\alpha$ triplet is now well fitted. So is the \nex~Ly$\alpha$. The CX contributes about a half of the thermal emission flux in the 7--30 \AA\ range.

Accounting for the CX results in considerable changes in the characterization of the thermal and chemical properties of the plasma  (Table ~\ref{tab:par}). The best-fit temperature $\sim 0.39$ keV in the {\it APEC+ACX} model is significantly higher than 0.25~keV in the fiducial {\it APEC} only fit. This is because CX tends to reduce the ionization level of the ions and emit softer line emission than the plasma itself. The accounting for the line emission by the CX via the inclusion of the {\it ACX} model component also reduces the need for high metal abundances. The kT $\sim 0.39$ keV plasma contributes little to the \ovii~line, but nearly dominates those highly ionized lines from \mgxi\ and  \fexvii. The emissivity of  Ne-like \fexvii\ also increases several times with the increased temperature, causing the decrease of the fitted iron abundance. The CX contributes little to \fexvii~lines, which need Fe$^{17+}$ ions that occupy a small ion fraction in the plasma. The abundance reduction for other elements are due to the large CX contribution to the relevant line emission: e.g., accounting for the majority of the \ovii~{\it f} line, producing the high $G$ ratio, as well as significant fractions of the  \oviii~Ly$\alpha$ to Ly$\delta$ lines. The \neix~{\it f} line is also enhanced by the CX, although the limited spectral quality of the data does not allow for a meaningful constraint on the $G$ ratio of the \neix~triplet. As a result, the  {\it APEC+ACX} model fit requires only solar-like abundances for the elements, except for the iron (Table~\ref{tab:par}). 

We next check how a temperature distribution of the plasma, which should be more realistic \citep{Wang21}, may affect the results. A simple extension from the single-temperature assumption is the use of the plasma model {\it vltnd} instead of {\it APEC}. This model adopts a log-normal temperature distribution of $x={\rm ln} T$  for the plasma emission measure and  has the mean  $\bar{x}$ and the dispersion $\sigma_x$  as two fitting parameters. Adopting {\it vltnd}, we also fit the temperature parameter in the {\it ACX} model to characterize the thermal property of ions undergoing CX.  Figure~\ref{fig:apeccx}c shows the best-fit  {\it vltnd+ACX} model. The quality of the fit is hardly changed because the fitted $\sigma_x$ is small, while the fitted  $\bar{T}$ and  metal abundances are also very similar to the best-fit values obtained in the above {\it APEC+ACX} model. The temperature of the CX ions is a bit less than 0.25 keV, which decreases the CX contribution to one-third of the plasma emission flux. So broadly speaking, the {\it APEC+ACX} model represents a simple, self-consistent characterization of the plasma  emission.

\subsection{\chandra/ACIS spectral analysis and results}	

\begin{figure}
\centering
      \includegraphics[angle=0,width=0.85\linewidth]{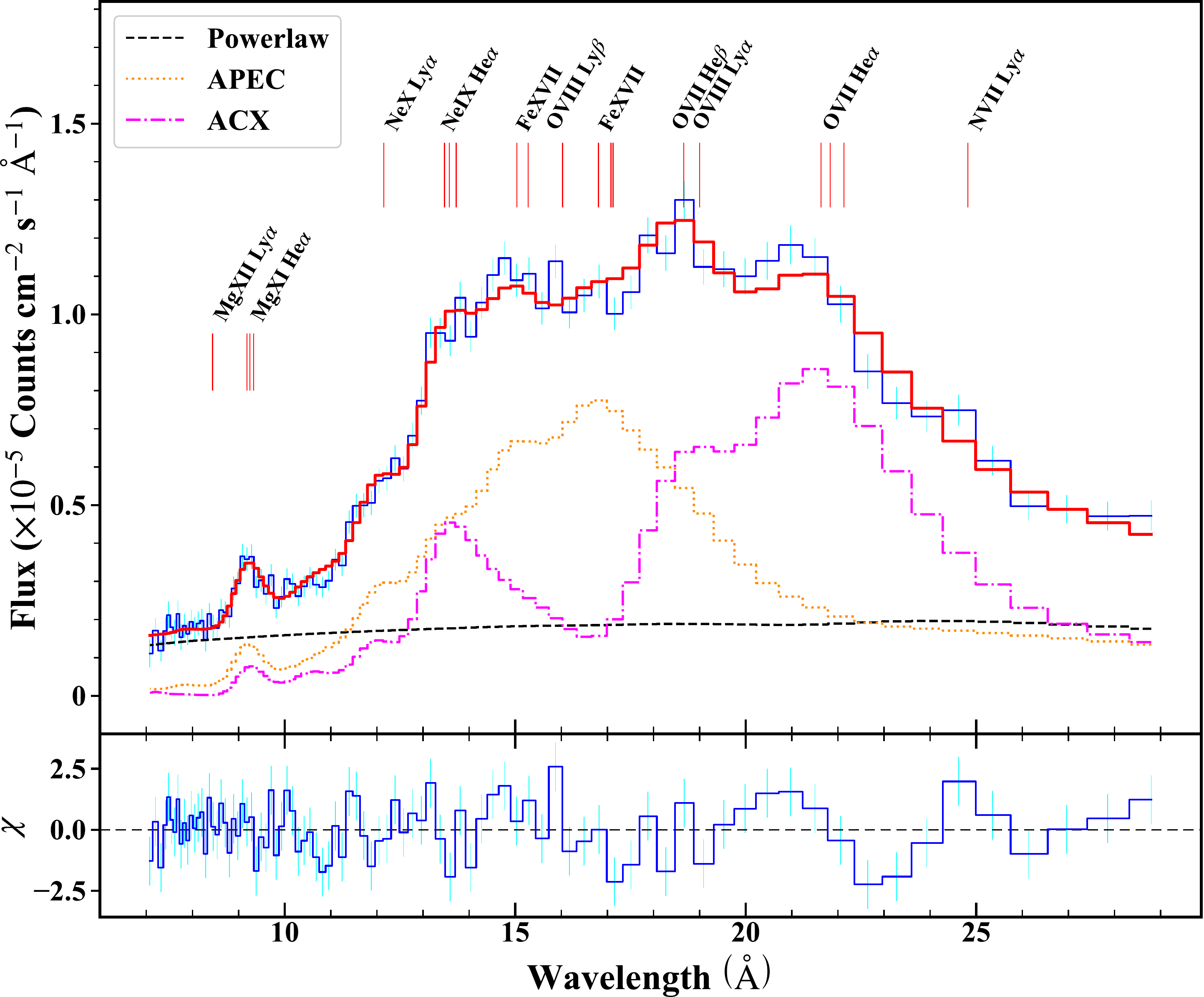} 
      \includegraphics[angle=0,width=0.85\linewidth]{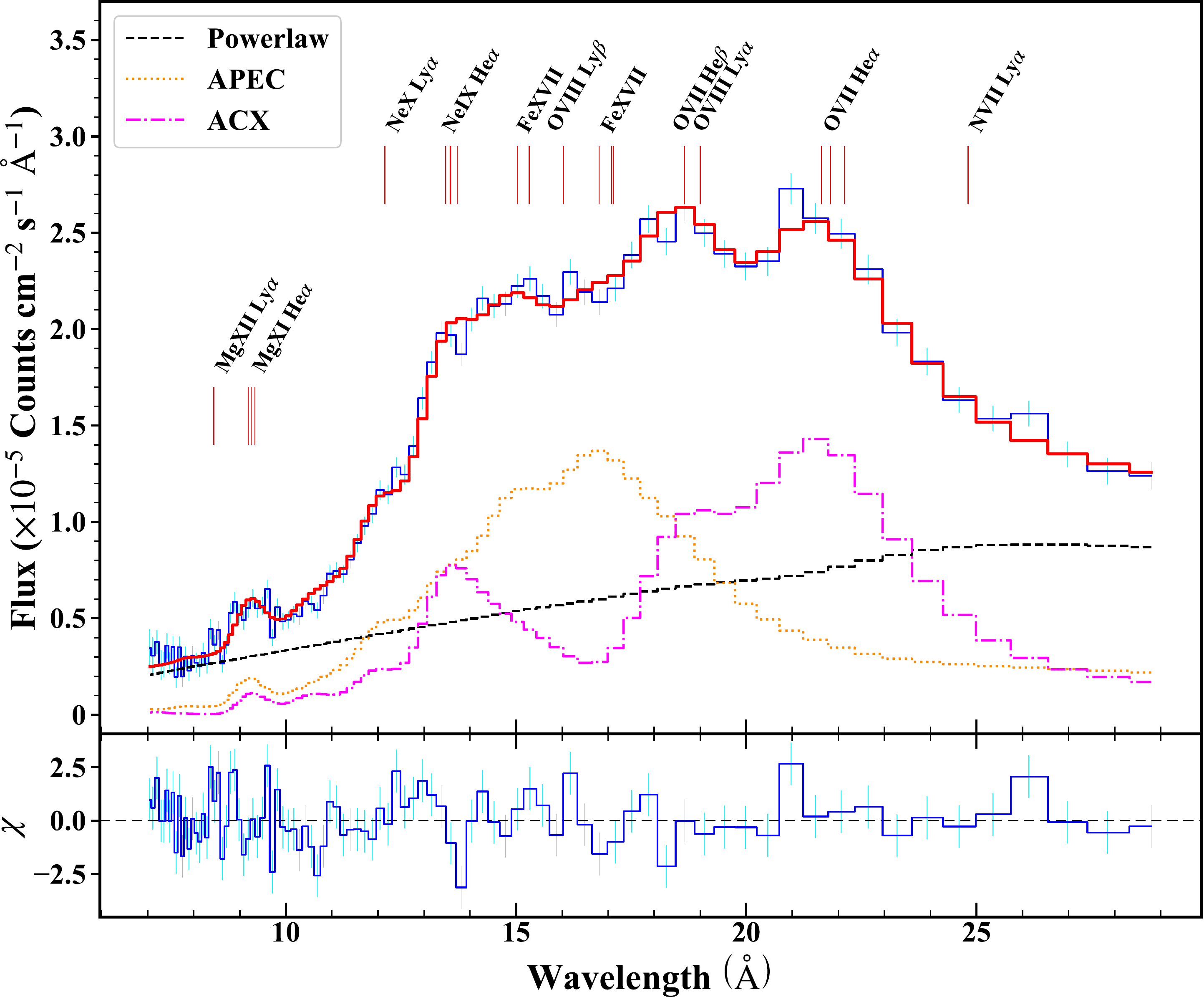} 
\caption{ ACIS-S spectra of the NHS and the galactic disk of M51 (upper and lower panels) in comparison with the best-fit thermal plasma plus CX  models. }
\label{fig:chandraspec} 
\end{figure}

Similarly, we analyze the \chandra/ACIS~spectra extracted from the NHS region and the M51 disk region without NHS, using various combinations of the CIE plasma and CX models. We find that the two-temperature {\it APEC} CIE plasma gives a reasonable characterization of the overall spectral shapes. The fitted temperatures are about 0.2 and 0.5 keV, similar to the results  (0.24 and 0.64 keV) from the analysis of  \xmm/EPIC spectral data on M51 \citep{Owen09}. The quality of the fits is not as good for the log-normal temperature model with one fewer fitting parameter. Nevertheless, the high \ovii~$G$ ratio from the RGS data rejects the scenario of a combination of thermal components with different temperatures. The simple {\it APEC+ACX} model fits the spectra well (Figure~\ref{fig:chandraspec}). The fits give similar spectral parameters for the two regions: their temperatures are both $\sim$0.34 keV, while the metal abundances are consistent with being solar within a factor of 2 (Table~\ref{tab:par}). The thermal emission peaks around 17 \AA\ and contributes mainly to \fexvii~lines, whereas the CX emission largely accounts for the double peaks at 14 \AA~and 22 \AA, as in the {\it APEC+ACX} fit to the RGS spectra of the NHS.

The high spatial resolution of the ACIS data, together with the consistent {\it APEC+ACX} model characterization of the spectral properties, allows us to improve the estimation of the CX contributions. In the 7--30 \AA\ range, the flux contribution from the CX is even higher than 50\% of the diffuse emission, either in the NHS region or in the disk region without NHS (Table~\ref{tab:par}). The areas of the two extraction regions of the \chandra~spectra are 5.45 and 19.78 arcmin$^2$, respectively. Therefore, the derived surface densities of either the thermal or the CX emission in the NHS region is about twice that in the disk region without NHS. 

\subsection{Spatial distributions of the oxygen line emission}
\label{sec:mapping}

\begin{figure*}[htbp] 
 \centering
       \includegraphics[angle=0,width=0.65\textwidth]{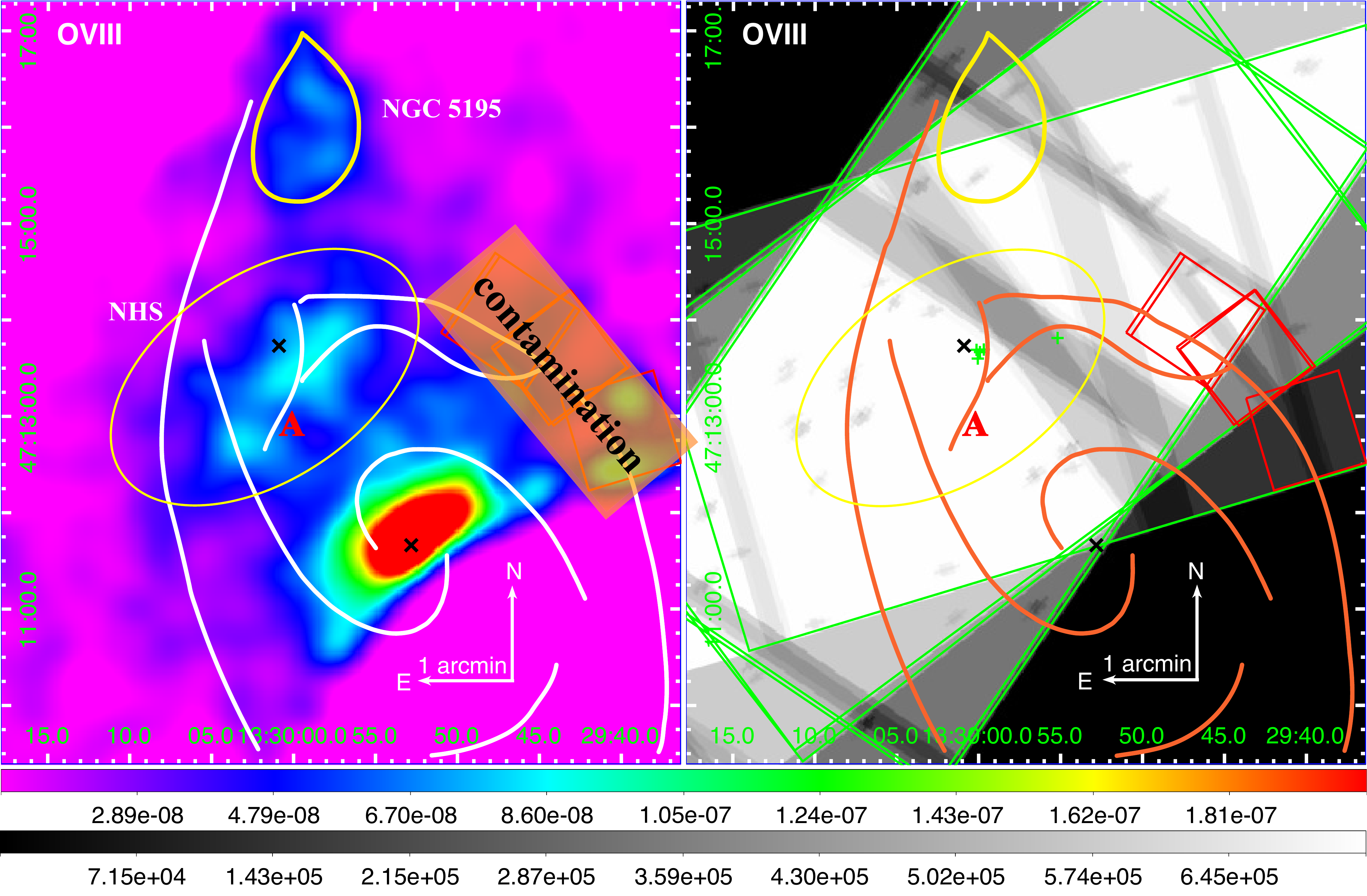}
       \includegraphics[angle=0,width=0.66\textwidth]{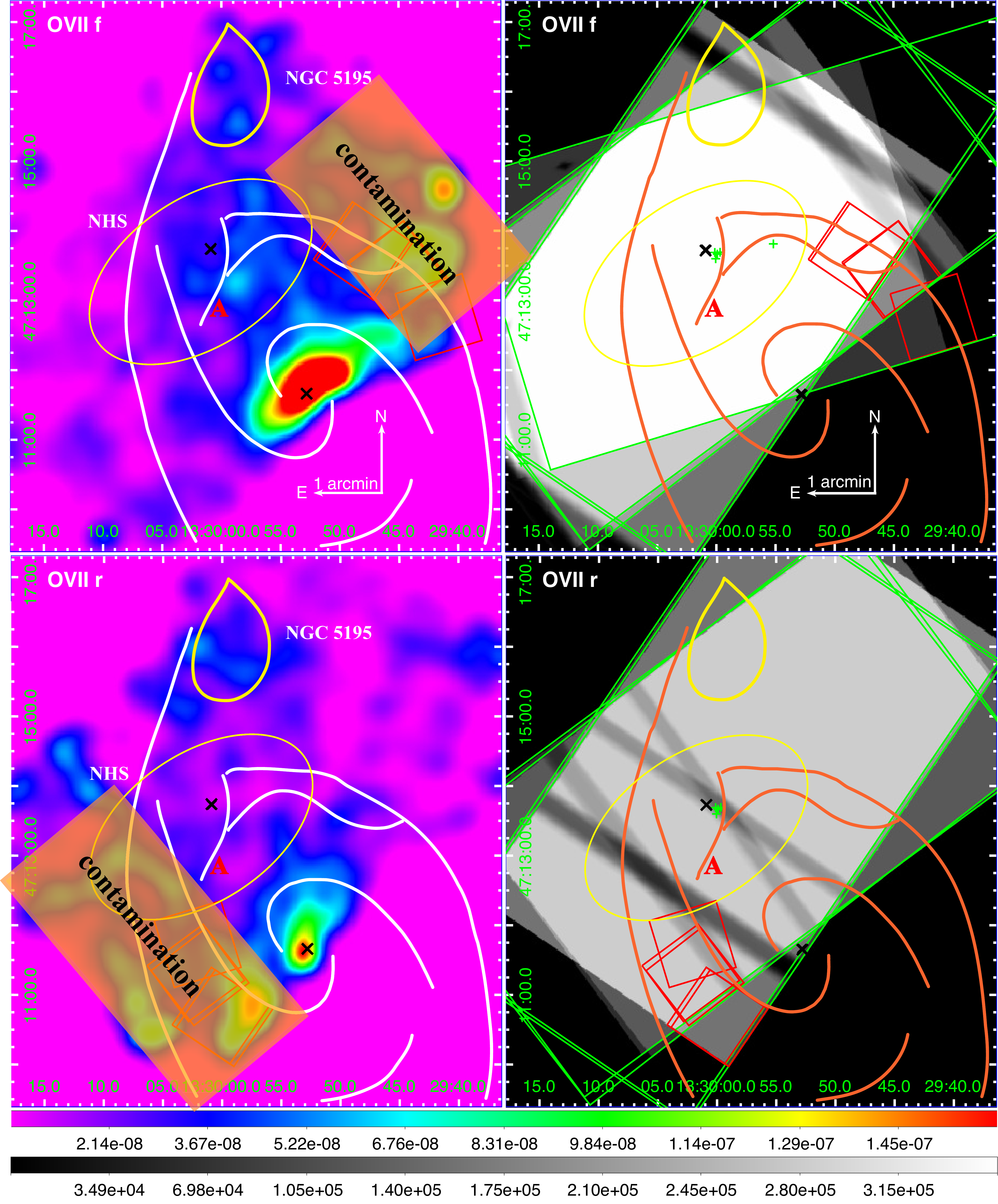}
 \caption{
Oxygen line intensity maps (left panels, in units of $\rm photon\,s^{-1}\,arcsec^{-2}$) and their corresponding RGS exposure maps (right panels, in units of seconds) of M51.
The morphology of the line intensity distribution in the galaxy's core  is distorted because of its location at edges of the RGS field of view. 
The smaller red boxes ($1'\times1'$) mark the regions that may be contaminated by the \ovii~K$\beta$ or \ovii~{\it i} line emission from the galactic core.
These contaminated regions, together with noisy regions due to obvious chip gaps, are further masked with the large orange rectangles.
The rest are the same as in Figure~\ref{fig:regions}. }
\label{fig:linemap}
\end{figure*}

\begin{figure*}[tp] 
 \centering
       \includegraphics[width=0.64\textwidth]{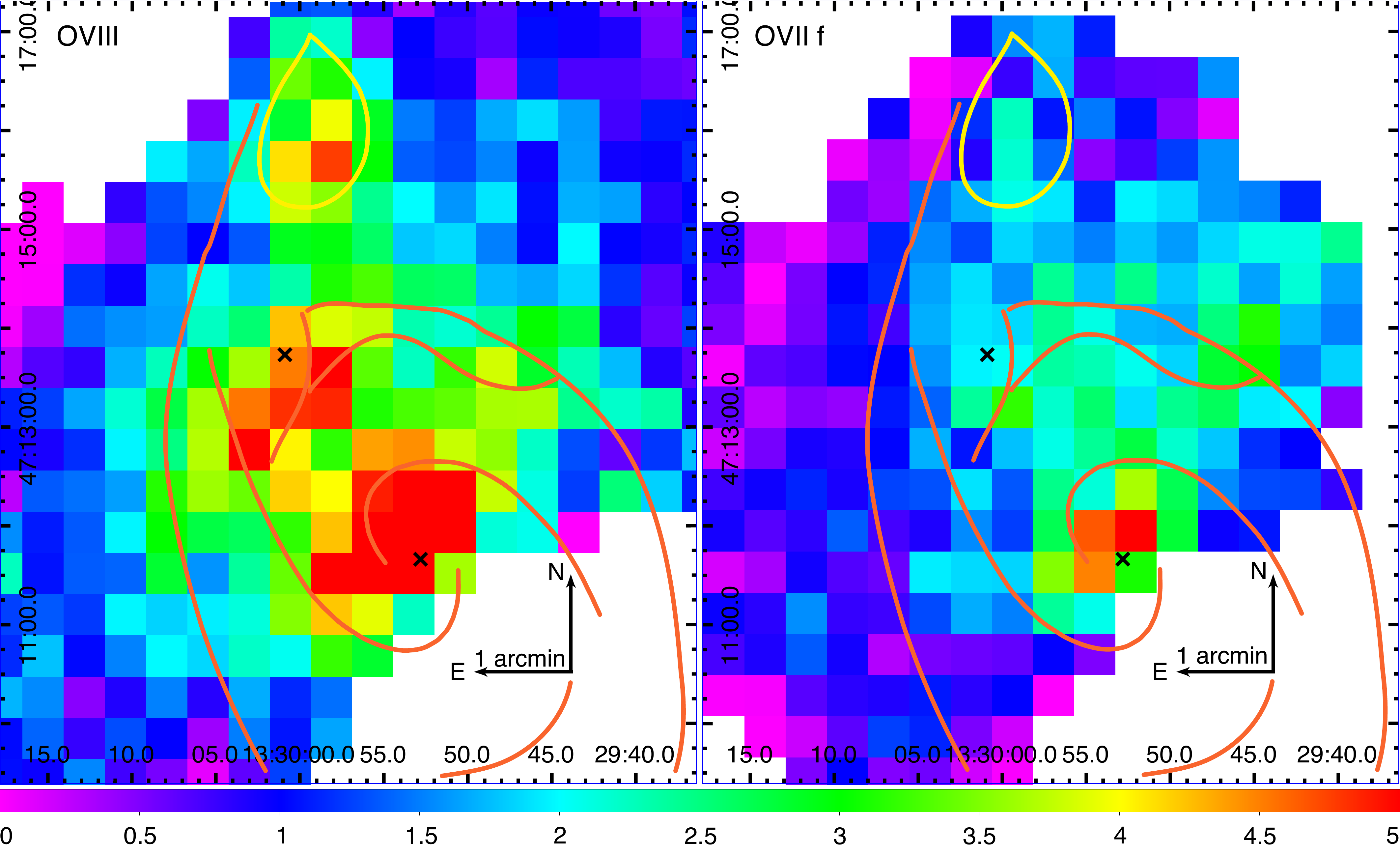}  
 \caption{S/N ratio maps of the \oviii~Ly$\alpha$ and the \ovii~He$\alpha$~{\it f}. The bin size is 25$''$, similar to that of the converted spatial resolution of RGS.}
\label{fig:SNr} 
\end{figure*}

Figure~\ref{fig:linemap} presents the line emission maps, while Figure~\ref{fig:SNr} shows the S/N maps for the two stronger lines. The constructions of these maps are detailed in \ref{sec:procedure}. Most features seen in the line emission maps at  the spatial resolution of $\sim0.5'$ seem reliable, especially in places where the S/N ratio is greater than $\sim 3$.  The counting statistics of the \ovii~line maps are not as good as those of the \oviii~map; the exposure time of the \ovii~lines is only about half of the \oviii. In the following, we compare the line RGS maps with the \chandra/ACIS diffuse soft X-ray emission image  (Figure~\ref{fig:regions}b), focusing on a few key features. 

The \oviii~map and the \chandra\ image look broadly similar. The M51 core region is prominent in the \oviii~emission, though its intensity morphology in the map is distorted because of the edge effect of the RGS CCDs. The line intensity distributions in both NGC 5195 and the NHS regions are consistent with those seen in the \chandra\ image, suggesting that the RGS reconstructed line map represents the spatial distribution of the diffuse X-ray emission well. 

At the M51 core, both \ovii~{\it r} and {\it f} lines are prominent, although the latter is significantly stronger than the former, consistent with the result from the RGS spectral analysis  \citep{Yang20}. In the NGC 5195 region,  the {\it r} line is generally brighter than the {\it f} line; the latter appears stronger at the southern end, where an arc-like structure is observed in the diffuse soft X-ray emission and is believed to be due to photoionization by an early AGN of the galaxy \citep{Schlegel16}. We find that the RGS spectrum of this structure does show a higher  \ovii~$G$ ratio than that of the entire NGC 5195 region, consistent with the AGN ionization scenario.

In the NHS region, the overall \ovii~{\it f} emission is obviously stronger than the {\it r} line emission, which is consistent with the high $G$ ratio obtained from the RGS spectral fitting. The region shows up as a local enhancement in the S/N maps, indicating a reliable detection of the  \oviii\ and \ovii~{\it f}  lines. We find that the \oviii\ and \ovii\ $f$ line fluxes estimated from the reconstructed RGS line maps, after accounting for the RGS effective areas at the corresponding wavelengths, are well consistent with those from the best-fit {\it APEC+ACX} model of the ACIS spectrum extracted from the same region.

Although the limited counting statistics of the RGS data do not allow for a reliable 2D structure determination of the line emission in the region, the distribution of \ovii~{\it f} appears to be different from that of \oviii\ or the diffuse soft X-ray emission seen in the \chandra\ image. The \ovii~{\it f} emission appears relatively brighter on the west side of the \hi\ segment ``A'',  consistent with the slight offset of the \ovii~{\it f} line peak in the RGS spectra to the blue side of the model line (Figure~\ref{fig:apeccx}b or c). In contrast, the \oviii~emission resembles the diffuse soft X-ray emission in Figure~\ref{fig:regions}b.

\section{Discussion}
\label{sec:discuss}

The above results demonstrate the apparent success of the thermal plasma plus CX modeling of the line emission from the M51 disk and further show differential distributions of the thermal and CX contributions. In the following, we further explore how the hot plasma interacts with the surrounding cool gas.

The diffuse soft X-ray enhancement observed at the NHS is clearly a result of the energetic mechanical feedback expected from massive stars. Their locations are closely aligned with the \hi\ segment \citep{Egusa17}, whereas the X-ray enhancement is lopsided toward the east  (Figure 1b). This structure is a natural manifestation of the high-pressure diffuse hot plasma, heated by the stellar feedback, expanding preferentially toward interarm regions of relatively low ISM density. This expansion may also explain the apparent bifurcation of the grand spiral arm at the location of the NHS, as seen in the \hi\ map (Figure 1c). The CX is expected to be more significant in places with abundant \hi, hence along the segment. 

With the thermal plasma+CX modeling for the diffuse soft X-ray emission, we may now estimate the physical properties of the plasma and the effective area ($A$ ) of its interface with cool gas. The \chandra~image (Figure~\ref{fig:regions}a) shows many fine structures in the NHS region caused by the absorption of dust lanes. So it is reasonable to assume that the plasma is largely contained in a thick galactic disk with an effective half-height of 0.5 kpc. We infer from the `norm' value of the {\sl APEC} component (Table~\ref{tab:par}) the mean density as $n \approx 0.0066\,{\rm cm^{-3}}$ in the NHS region or  $0.0046\,{\rm cm^{-3}}$ in the disk region without NHS. Similarly, we may infer the effective area from the `norm' of the ACX component, as defined in the note to Table~\ref{tab:par}. In the definition, the volume of the interface layer $V_\mathrm{i}$ is approximately the product of  $A$ and the mean free path of hot ions,  $l = \left( \sigma n_\mathrm{d} \right)^{-1}$, where  the CX cross section is typically $\sigma \sim 3\times 10^{-15}$~cm$^{-2}$. Accordingly, the `norm' of ACX can then be expressed as
\begin{equation}
  \eta_\mathrm{vacx} = \frac{10^{-10}}{4\pi D^2} \int \frac{n_\mathrm{r}}{\sigma} \mathrm{d} A.
  \label{eq:vacx_norm_explicit}
\end{equation}
Taking $n_{r} \sim n = 0.0066$ or 0.0046 \cmcu, we  estimate $A = 1.9\times10^{45}\,\rm cm^2$ or $6.5\times10^{45}\,\rm cm^2$ for the NHS or the other portion of the galactic disk, respectively. Although the diffuse X-ray surface intensity is higher in the NHS than in the other portion of the disk, the ratio between the estimated $A$ value and the physical area of the spectral extraction region (with the $20^\circ$ inclination angle taken into account) hardly changes: 5.5 and 5.2 for the two regions, respectively. This probably indicates that the CX effectiveness does not change much across much of the galaxy.

The presence of the CX may indicate significant effect on the cooling of the hot plasma. The radiative cooling time of the tenuous plasma in the disk is on the order of $10^{9}$ yr. However, under the effective interplay between the hot plasma and the cold ISM, as revealed by the large CX interface area, the plasma in the disk plane may be well mixed with the cold gas and cool down much quicker. The CX emission itself is from the ionization energy of ions, and it contributes energy similar to that of the thermal emission, mainly in the UV and X-ray bands. A prominent coolant line of the CX is  \heii~Ly$\alpha$ line (303.8 \AA) in the UV band, whose photon flux could be 20 times stronger than the total radiating photons in the RGS band.

\section{Summary}
\label{sec:conclusion}
In this work, we have investigated the diffuse X-ray emission from the galactic disk of the star-forming galaxy M51. In particular, we have conducted a spectroscopic analysis of the emission from the NHS, an enhanced star-formation region off the galactic nucleus, based on the high spectral resolution data from \xmm/RGS. The same data have also allowed us to reconstruct 2D oxygen line maps, which cover much of the galactic disk . Furthermore, we have used all the available complementary \chandra/ACIS-S data to perform imaging and spectral analyses of the emission across the entire galactic disk. Our main results and conclusions are as follows:

1. The RGS spectrum of the NHS shows a high $G$ ratio of the \ovii~triplet ($\sim3.2$). The \ovii~{\it  f} emission appears significantly stronger than the {\it r}, which is also confirmed by the RGS \ovii~emission-line mapping. This high $G$ ratio is inconsistent with the interpretation of the line emission purely originating from CIE plasma, even when various possible temperature distributions are considered.

2. The oxygen line maps from the RGS data share similarities with the diffuse soft X-ray image obtained from the ACIS-S data. The \ovii~{\it f} emission in the companion galaxy NGC 5195 field is only prominent around a southern arc, suggesting the presence of an AGN photoionization remnant. In the NHS region, the \oviii~emission appears relatively bright to the east of the \hi\ arm segment ``A'', where the star formation is intense. In contrast, the \ovii~{\it f} line is more luminous on the west side of the segment where the \hi~gas is more abundant. 

3. The CX naturally explains both the high \ovii~$G$ ratio and the differential spatial distributions of the \oviii\ and \ovii\ {\it f} line intensities in the NHS region, because the CX tends to occur in and near star-forming regions where both hot and cool gases are abundant. In this scenario, the \ovii~{\it f} line preferentially traces the CX distribution, while the \oviii~emission is contributed by both the thermal plasma and the CX. The thermal emission from the hot plasma is expected to be more extended, as is the case in the NHS region. No other simple mechanism can simultaneously explain all the spectral and spatial characteristics observed in the region.

4. The spectral modeling of the ACIS data with the inclusion of the CX contribution further enables us to estimate the properties of the hot plasma in the M51 disk. The spectra of the plasma can be well characterized by a single {\it APEC} with a temperature $\sim$0.34 keV and solar-like metal abundances, which seem to be only weakly dependent of the soft X-ray surface intensity.

5. The CX contributes about a half of the diffuse X-ray emission in the 7--30 \AA~band, which again seems to weakly depend on the surface intensity. We estimate the effective interface area to be about five times the geometric area in the galactic disk, suggesting that the CX may represent a potentially important cooling mechanism for diffuse hot plasma in star-forming galaxies.

S.N.Z. acknowledges the support from the NSFC grant 11573070 and the China Scholarship Council. This work has made use of the data from \xmm\ and \chandra, and the fits image from ``The \hi\ Nearby Galaxy Survey.'' We thank the anonymous referee for the constructive comments and suggestions.

\bibliography{whirlpool}

\begin{thebibliography}{}
\expandafter\ifx\csname natexlab\endcsname\relax\def\natexlab#1{#1}\fi
\providecommand{\url}[1]{\href{#1}{#1}}
\providecommand{\dodoi}[1]{doi:~\href{http://doi.org/#1}{\nolinkurl{#1}}}
\providecommand{\doeprint}[1]{\href{http://ascl.net/#1}{\nolinkurl{http://ascl.net/#1}}}
\providecommand{\doarXiv}[1]{\href{https://arxiv.org/abs/#1}{\nolinkurl{https://arxiv.org/abs/#1}}}

\bibitem[{{Bauer} {et~al.}(2007){Bauer}, {Pietsch}, {Trinchieri},
  {Breitschwerdt}, {Ehle}, \& {Read}}]{Bauer07}
{Bauer}, M., {Pietsch}, W., {Trinchieri}, G., {et~al.} 2007, \aap, 467, 979,
  \dodoi{10.1051/0004-6361:20066340}

\bibitem[{{Bertin}(2010)}]{Bertin10}
{Bertin}, E. 2010, {SWarp: Resampling and Co-adding FITS Images Together}.
\newblock \doeprint{1010.068}

\bibitem[{{Breitschwerdt} \& {Schmutzler}(1999)}]{Breitschwerdt99}
{Breitschwerdt}, D., \& {Schmutzler}, T. 1999, \aap, 347, 650.
\newblock \doarXiv{astro-ph/9902268}

\bibitem[{{Calzetti} {et~al.}(2005){Calzetti}, {Kennicutt}, {Bianchi},
  {Thilker}, {Dale}, {Engelbracht}, {Leitherer}, {Meyer}, {Sosey}, {Mutchler},
  {Regan}, {Thornley}, {Armus}, {Bendo}, {Boissier}, {Boselli}, {Draine},
  {Gordon}, {Helou}, {Hollenbach}, {Kewley}, {Madore}, {Martin}, {Murphy},
  {Rieke}, {Rieke}, {Roussel}, {Sheth}, {Smith}, {Walter}, {White}, {Yi},
  {Scoville}, {Polletta}, \& {Lindler}}]{Calzetti05}
{Calzetti}, D., {Kennicutt}, Jr., R.~C., {Bianchi}, L., {et~al.} 2005, \apj,
  633, 871, \dodoi{10.1086/466518}

\bibitem[{{Cheng} {et~al.}(2021){Cheng}, {Wang}, \& {Lim}}]{Cheng21}
{Cheng}, Y., {Wang}, Q.~D., \& {Lim}, S. 2021, \mnras, 504, 1627,
  \dodoi{10.1093/mnras/stab1040}

\bibitem[{{Doane} {et~al.}(2004){Doane}, {Sanders}, {Wilcots}, \&
  {Juda}}]{Doane04}
{Doane}, N.~E., {Sanders}, W.~T., {Wilcots}, E.~M., \& {Juda}, M. 2004, \aj,
  128, 2712, \dodoi{10.1086/425627}

\bibitem[{{Dobbs} {et~al.}(2010){Dobbs}, {Theis}, {Pringle}, \&
  {Bate}}]{Dobbs10}
{Dobbs}, C.~L., {Theis}, C., {Pringle}, J.~E., \& {Bate}, M.~R. 2010, \mnras,
  403, 625, \dodoi{10.1111/j.1365-2966.2009.16161.x}

\bibitem[{{Egusa} {et~al.}(2017){Egusa}, {Mentuch Cooper}, {Koda}, \&
  {Baba}}]{Egusa17}
{Egusa}, F., {Mentuch Cooper}, E., {Koda}, J., \& {Baba}, J. 2017, \mnras, 465,
  460, \dodoi{10.1093/mnras/stw2710}

\bibitem[{{Hodges-Kluck} {et~al.}(2018){Hodges-Kluck}, {Bregman}, \&
  {Li}}]{Hodges-Kluck18}
{Hodges-Kluck}, E.~J., {Bregman}, J.~N., \& {Li}, J.-t. 2018, \apj, 866, 126,
  \dodoi{10.3847/1538-4357/aae38a}

\bibitem[{{Kalberla} {et~al.}(2005){Kalberla}, {Burton}, {Hartmann}, {Arnal},
  {Bajaja}, {Morras}, \& {P{\"o}ppel}}]{Kalberla05}
{Kalberla}, P.~M.~W., {Burton}, W.~B., {Hartmann}, D., {et~al.} 2005, \aap,
  440, 775, \dodoi{10.1051/0004-6361:20041864}

\bibitem[{{Kaleida} \& {Scowen}(2010)}]{Kaleida10}
{Kaleida}, C., \& {Scowen}, P.~A. 2010, \aj, 140, 379,
  \dodoi{10.1088/0004-6256/140/2/379}

\bibitem[{{Kuntz} \& {Snowden}(2010)}]{Kuntz10}
{Kuntz}, K.~D., \& {Snowden}, S.~L. 2010, \apjs, 188, 46,
  \dodoi{10.1088/0067-0049/188/1/46}

\bibitem[{{Li} \& {Wang}(2013)}]{Li13}
{Li}, J.-T., \& {Wang}, Q.~D. 2013, \mnras, 428, 2085,
  \dodoi{10.1093/mnras/sts183}

\bibitem[{{McKee} \& {Ostriker}(1977)}]{McKee77}
{McKee}, C.~F., \& {Ostriker}, J.~P. 1977, \apj, 218, 148,
  \dodoi{10.1086/155667}

\bibitem[{{McQuinn} {et~al.}(2016){McQuinn}, {Skillman}, {Dolphin}, {Berg}, \&
  {Kennicutt}}]{McQuinn16}
{McQuinn}, K.~B.~W., {Skillman}, E.~D., {Dolphin}, A.~E., {Berg}, D., \&
  {Kennicutt}, R. 2016, \apj, 826, 21, \dodoi{10.3847/0004-637X/826/1/21}

\bibitem[{{Mineo} {et~al.}(2012){Mineo}, {Gilfanov}, \& {Sunyaev}}]{Mineo12}
{Mineo}, S., {Gilfanov}, M., \& {Sunyaev}, R. 2012, \mnras, 426, 1870,
  \dodoi{10.1111/j.1365-2966.2012.21831.x}

\bibitem[{{Mullen} {et~al.}(2017){Mullen}, {Cumbee}, {Lyons}, {Gu}, {Kaastra},
  {Shelton}, \& {Stancil}}]{Mullen2017}
{Mullen}, P.~D., {Cumbee}, R.~S., {Lyons}, D., {et~al.} 2017, \apj, 844, 7,
  \dodoi{10.3847/1538-4357/aa7752}

\bibitem[{{Owen} \& {Warwick}(2009)}]{Owen09}
{Owen}, R.~A., \& {Warwick}, R.~S. 2009, \mnras, 394, 1741,
  \dodoi{10.1111/j.1365-2966.2009.14464.x}

\bibitem[{{Schlegel} {et~al.}(2016){Schlegel}, {Jones}, {Machacek}, \&
  {Vega}}]{Schlegel16}
{Schlegel}, E.~M., {Jones}, C., {Machacek}, M., \& {Vega}, L.~D. 2016, \apj,
  823, 75, \dodoi{10.3847/0004-637X/823/2/75}

\bibitem[{{Simmonds} {et~al.}(2021){Simmonds}, {Schaerer}, \&
  {Verhamme}}]{Simmonds21}
{Simmonds}, C., {Schaerer}, D., \& {Verhamme}, A. 2021, \aap, 656, A127,
  \dodoi{10.1051/0004-6361/202141856}

\bibitem[{{Smith} {et~al.}(2012){Smith}, {Foster}, \& {Brickhouse}}]{Smith12}
{Smith}, R.~K., {Foster}, A.~R., \& {Brickhouse}, N.~S. 2012, Astronomische
  Nachrichten, 333, 301, \dodoi{10.1002/asna.201211673}

\bibitem[{{Terashima} \& {Wilson}(2004)}]{Terashima04}
{Terashima}, Y., \& {Wilson}, A.~S. 2004, \apj, 601, 735,
  \dodoi{10.1086/380505}

\bibitem[{{Thilker} {et~al.}(2000){Thilker}, {Braun}, \&
  {Walterbos}}]{Thilker00}
{Thilker}, D.~A., {Braun}, R., \& {Walterbos}, R. A.~M. 2000, \aj, 120, 3070,
  \dodoi{10.1086/316852}

\bibitem[{{van der Heyden} {et~al.}(2003){van der Heyden}, {Bleeker},
  {Kaastra}, \& {Vink}}]{Heyden03}
{van der Heyden}, K.~J., {Bleeker}, J.~A.~M., {Kaastra}, J.~S., \& {Vink}, J.
  2003, \aap, 406, 141, \dodoi{10.1051/0004-6361:20030658}

\bibitem[{{Walter} {et~al.}(2008){Walter}, {Brinks}, {de Blok}, {Bigiel},
  {Kennicutt}, {Thornley}, \& {Leroy}}]{Walter08}
{Walter}, F., {Brinks}, E., {de Blok}, W.~J.~G., {et~al.} 2008, \aj, 136, 2563,
  \dodoi{10.1088/0004-6256/136/6/2563}

\bibitem[{{Wang} \& {Liu}(2012)}]{Wang12}
{Wang}, Q.~D., \& {Liu}, J. 2012, Astronomische Nachrichten, 333, 373,
  \dodoi{10.1002/asna.201211657}

\bibitem[{{Wang} {et~al.}(2021){Wang}, {Zeng}, {Bogd{\'a}n}, \& {Ji}}]{Wang21}
{Wang}, Q.~D., {Zeng}, Y., {Bogd{\'a}n}, {\'A}., \& {Ji}, L. 2021, \mnras, 508,
  6155, \dodoi{10.1093/mnras/stab2997}

\bibitem[{{Warwick} {et~al.}(2007){Warwick}, {Jenkins}, {Read}, {Roberts}, \&
  {Owen}}]{Warwick07}
{Warwick}, R.~S., {Jenkins}, L.~P., {Read}, A.~M., {Roberts}, T.~P., \& {Owen},
  R.~A. 2007, \mnras, 376, 1611, \dodoi{10.1111/j.1365-2966.2007.11571.x}

\bibitem[{{Yang} {et~al.}(2020){Yang}, {Zhang}, \& {Ji}}]{Yang20}
{Yang}, H., {Zhang}, S., \& {Ji}, L. 2020, \apj, 894, 22,
  \dodoi{10.3847/1538-4357/ab80c9}

\bibitem[{{Zhang} {et~al.}(2019){Zhang}, {Wang}, {Foster}, {Sun}, {Li}, \&
  {Ji}}]{Zhang19}
{Zhang}, S., {Wang}, Q.~D., {Foster}, A.~R., {et~al.} 2019, \apj, 885, 157,
  \dodoi{10.3847/1538-4357/ab4a0f}

\bibitem[{{Zhang} {et~al.}(2014){Zhang}, {Wang}, {Ji}, {Smith}, {Foster}, \&
  {Zhou}}]{Zhang14}
{Zhang}, S., {Wang}, Q.~D., {Ji}, L., {et~al.} 2014, \apj, 794, 61,
  \dodoi{10.1088/0004-637X/794/1/61}

\end{thebibliography}
\label{lastpage}

\begin{appendix}
\section{Procedure for Using RGS Data to Map Line Emission}
\label{sec:procedure}
The procedure is designed for extracting the maps of the line emission from the key ions, \ovii~and \oviii, expected for diffuse hot plasma in the ISM, based on the assumption that the spatial broadening dominates while the dynamic broadening is negligible. The \ovii~and \oviii~lines, in particular, are chosen for several reasons: 
1) the \ovii~{\it f} line and the \oviii~Ly$\alpha$ lines may be dominated by different processes, e.g. thermal, CX, photoinization, and/or recombination processes;
2) these lines are strong enough in the RGS spectra; 
3) the lines are relatively isolated.
The input files are generated by the SAS pipeline ``{\tt rgsproc}'' for the first order grating data, as described in the RGS data reduction in Sec. \ref{sec:data}.

The usefulness of our constructed line emission maps are limited by the spatial extendedness of M51. Its X-ray angular diameter  is about 5$'$, corresponding to 0.69 \AA. For the \oviii~extraction, we use the wavelength interval 18--20 \AA, which fully covers galaxy in the dispersion direction. The situation for the \ovii~{\it f} line and {\it r} is more complicated, because they are separated by only 0.5 \AA. We thus use the narrower intervals: 21.7--22.7 \AA\  for the {\it f} line and  21--22 \AA\  for the {\it r} line. Again ObsID 0852030101 is not used for producing the {\it r} line map, due to the bad column. The overlapping of these intervals results in the line mapping confusion between the two lines. But because the five observations have similar dispersion directions, this confusion is largely limited to the west part of the {\it f} map and the east part of the {\it r} map.

\begin{figure}
 \centering
       \includegraphics[angle=0,width=1\linewidth]{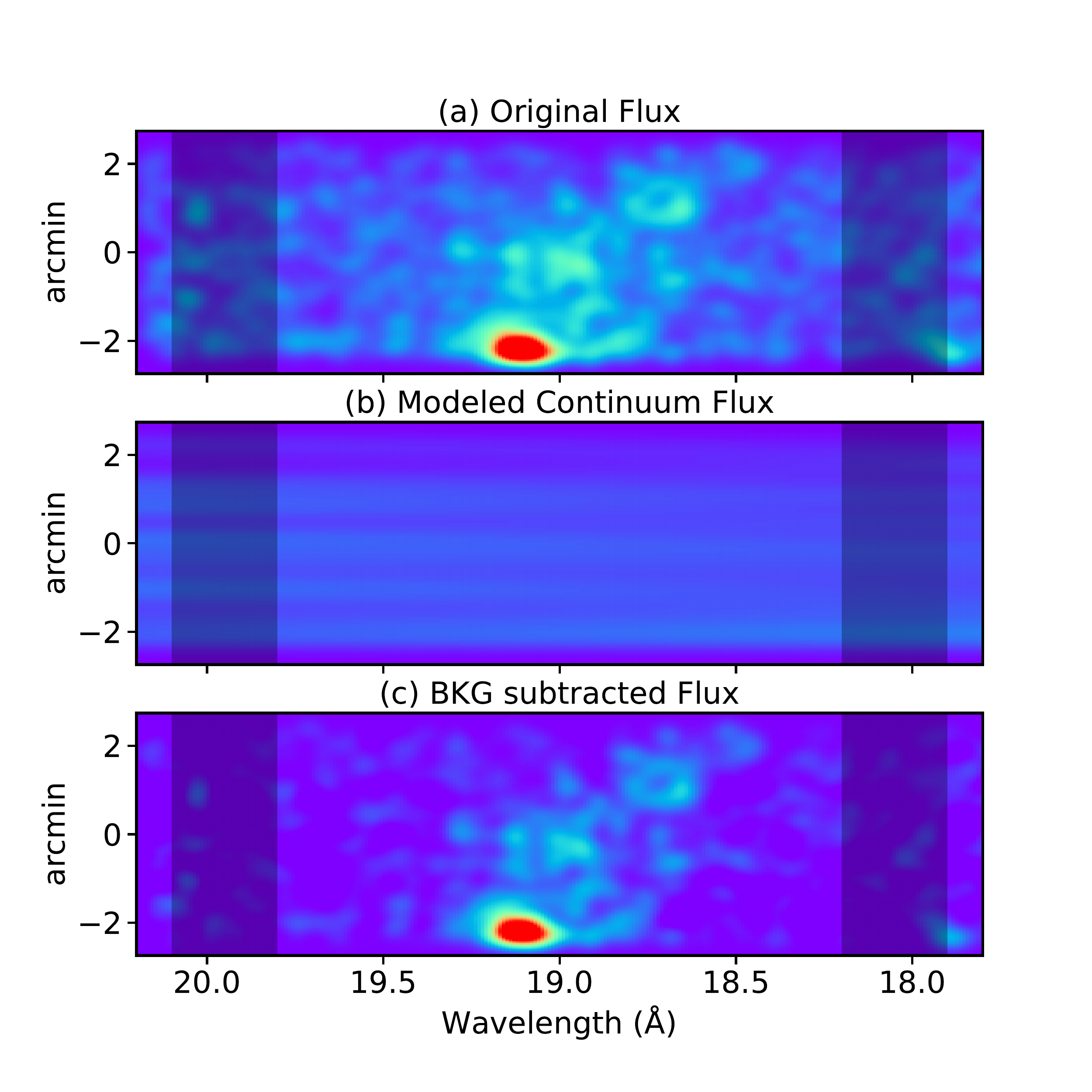}  
 \caption{Illustration of the \oviii~map construction for the RGS observation (ID: 0824450901): (a) the count intensity map (without the continuum removal), smoothed with a Gaussian of size comparable to the resolution of the data; (b) the continuum intensity map interpolated from two off-line bands marked by the two shaded regions; (c) the line flux map after the continuum subtraction.}
\label{fig:extraction} 
\end{figure}

For each observation, we first obtain the count and exposure images for both RGS1 and RGS2. The pixel size of the images inherits from the CCD pixel, 0.01 \AA\ $\times\,2\farcs22$. In the XDSP direction, the effective area around the CCD edges decreases dramatically, and thus we cut it within $\pm2\farcm2$. The telescope mirror vignetting effect is less than 5\%--8\% in the 2$'$--3$'$ off-axis angle range and is thus not corrected. In the dispersion direction, we select the data within the chosen wavelength intervals, as well as in the corresponding CCD energy range within $\pm$160 eV of each line. With the RGS exposure information directly given by the pipeline, we construct the exposure image accordingly. We can construct a count intensity image by dividing the count image with the exposure image,  as shown in Figure~\ref{fig:extraction}a. However, such an image also contains nonline emission dominated by the continuum contribution from pointlike sources, as well as various background components. We then estimate the contribution by interpolating the intensities in the off-line bands. For the \oviii~line, we choose 17.9--18.2 \AA\ and 19.8--20.1 \AA, as shown in Figure~\ref{fig:extraction}. The interpolation is done, row by row, between the mean intensities of the two bands, linearly in the log-log space (equivalent to assuming a power law for the continuum spectrum). By subtracting the resultant continuum image (Figure~\ref{fig:extraction}b), we obtain the \oviii\ line intensity image (Figure~\ref{fig:extraction}c).

To construct the line map from the multiple RGS observations, we first merge their count, continuum, and exposure images into the corresponding maps. This way, the counting statistical information of the data is preserved for subsequent analysis. To do so, we first regulate the pixel size of the images from individual observations to $4\farcs32\times2\farcs22$ and then merge them into the corresponding maps with a refined pixel size of $1''\times1''$ for better visualization, using the software SWarp \citep{Bertin10}. Finally, by dividing the combined total count map minus the corresponding continuum map with the combined exposure map, we obtain the combined line intensity map. Figure~\ref{fig:linemap} shows the map smoothed with a Gaussian of $0\farcm5$ FWHM.

We construct the intensity maps for the \ovii~lines, following the same procedure. The continuum maps are interpolated from the bands 20.23--20.55 \AA\ and 22.8--23.3 \AA. While the \oviii~maps are constructed from RGS1 and RGS2,  the \ovii~maps are from RGS1 only.

To assess the significance of features in a line intensity map, we further construct a corresponding S/N map. While this S/N map can simply be obtained from the ratio: $\mathrm{line\_count\_map}/\sqrt{\mathrm{total\_count\_map}}$, we rebin the count maps to 
a bin size of 25$'' \times 25''$, comparable to the RGS spatial resolution, to increase the counting statistics sufficiently (Figure~\ref{fig:SNr}).

\section{Consideration of alternative explanations for the large $G$ ratio observed in the RGS spectra}
\label{sec:explanations}

In the main text, we regard the large $G$ ratio of the \ovii~He$\alpha$ triplet  as the evidence for the CX contribution to the RGS spectra of the NHS. Here we consider various potential alternative explanations, which to the end are not favored. First, some rare discrete sources (e.g., supersoft ultraluminous sources) can show strong emission lines, which could contaminate the calculation of the $G$ ratio; but no such source is found inside the RGS extraction region \citep{Terashima04}.

Second, we  investigate other competing scenarios with the possibility to have a high \ovii~$G$ ratio, including the photoionization or the nonequilibrium ionization (NEI) processes that can increase the {\it f} line emission, and the resonance scattering (RS) process that has a chance to reduce the {\it r} line emission. 

For the photoionization scenario, the production of a high \ovii~$G$ ratio is possible through the recombination process, but it requires a powerful ionizing source such as a current or a past AGN to photoionize a plasma to form O$^{7+}$ or O$^{8+}$ ions. However, the NHS is too far away ($>5$ kpc) from the nuclear center of either M51 or NGC 5195. Even a bright AGN (e.g. $10^{46}\,\rm erg\,s^{-1}$) cannot photoionize the plasma in the NHS to such a highly ionized state, especially after the strong absorption by the arms in the disk. In situ, some ultra-luminous X-ray sources can also ionize surrounding materials and produce diffuse nebular emission \citep[e.g.,][]{Simmonds21}. However, in the NHS region, only two sources are more luminous than $10^{39}\,\rm{ergs\,s^{-1}}$ \citep{Terashima04}, and their contribution to the \ovii~{\it f} line is negligible compared to the estimated value of $1.25\times10^{-5}$ \phcm\ in the NHS region according to the RGS map.

The NEI scenario includes  ``overionized'' and  ``underionized'' cases. If the hot plasma has cooled substantially due to quick adiabatic expansion to an electron temperature lower than $10^6$ K, while the ions are still in the ``overionized'' state, the recombining spectrum could show a high \ovii~$G$ ratio. In an adiabatic model of the winds from spiral galaxies, cooling becomes noticeable at distances $>$3 kpc from the disk \citep{Breitschwerdt99}, though it takes tens of million years for the hot plasma to arrive there. However, the density of the hot plasma in the disk is as low as $n=0.005\,{\rm cm^{-3}}$ (see Sec.~\ref{sec:discuss}), which may decrease to one-tenth of this value after the expansion. Plus the low emissivity in the recombining process, a large volume of plasma is required to produce the observed \ovii~{\it f} line emission, for example with a height of a few hundred kiloparsecs over the disk. In this case, the line emission should be very diffuse over the disk, but in our map the \ovii~{\it f} enhancement is still mainly confined in the NHS region.

In the case of the ``underionized'' nonequilibrium, the lithium-like oxygen ions may produce the \ovii~{\it f} line emission. In some special case such as shock heated gas in supernova remnants, where the hydrogen has already been fully ionized and the ionization of heavier elements reaches only intermediate levels, the high \ovii~$G$ ratio may exist. However, the lithium-like oxygen ions will turn into the helium-like ions in a short timescale, and thus have a low possibility to produce the observed flux of the \ovii~{\it f} line. In addition, under this situation, the fitted temperature according to metal lines of the moderately ionized ions should be lower than 0.1 keV. But there are no hints of this, even according to the log-normal model.

Resonance scattering is another possible scenario to generate the high \ovii~$G$ ratio, since the \ovii~{\it r} line photons may be scattered to a more diffuse state and outside the spectral extraction region. However, the {\it r} line map that covers a large enough area is just generally weak in the NHS region compared to the {\it f} line map. Besides, this scenario requires a large column density of O$^{6+}$ ions over the disk, which is not achievable with the current best-fit model of the plasma.

We conclude that these competing scenarios are unlikely the reasons of the high \ovii~$G$ ratio in the NHS. The CX emission is generally preferred.

\end{appendix}

\end{document}